\documentclass[twocolumn]{aastex6}
\usepackage{natbib}
\usepackage{color}
\usepackage{hyperref} 

\def    \apjl  		{\rm {ApJL}}

\def    \apjl  		{\rm {ApJL}}

\def	\cm		{\,{\rm {cm}}}
\def	\K		{\,{\rm K}}
\def	\g		{\,{\rm {g}}}
\def	\mum	{\,{\mu \rm{m}}}

\def \bea {\begin{eqnarray}}
\def \ena {\end{eqnarray}}

\def	\ted	{{\tau_{\rm ed}}}
\def	\tH	{{\tau_{\rm H}}}

\def    \bmu    {{\hbox{\boldsym\char'026}}}	

\def    \bomega {{\hbox{\boldsym\char'041}}}	

\def	\cm	{\,{\rm cm}}

\def	\D	{{\rm D}}

\def	\ed	{{\rm ed}}

\def	\erg	{\,{\rm ergs}}

\def	\g	{\,{\rm g}}
\def	\gas	{\,{\rm gas}}

\def	\GHz	{\,{\rm GHz}}

\def	\H	{{\rm H}}

\def	\K	{{\rm K}}

\def	\s	{\,{\rm s}}

\def	\AU	{{\rm AU}}

\def    \gas     	{{\rm gas}}

\font\mib=cmmib10

\def\bomega{\hbox{\mib\char"21}}

\def\bmu{\hbox{\mib\char"16}}

\begin{document}
\shorttitle{Spinning dust from circumstellar disks}
\shortauthors{Hoang, Lan, Vinh, and Kim}
\title{Spinning dust emission from circumstellar disks and its role in excess microwave emission}
\author{Thiem Hoang}
\affil{Korea Astronomy and Space Science Institute, Daejeon 34055, South Korea; \href{mailto:thiemhoang@kasi.re.kr}{thiemhoang@kasi.re.kr}}
\affil{Korea University of Science and Technology, 217 Gajeong-ro, Yuseong-gu, Daejeon, 34113, South Korea}

\author{Nguyen-Quynh Lan}
\affil{Center for Astrophysics, Department of Physics, University of Notre Dame, Notre Dame, IN 46556; \href{mailto:lnguyen3@nd.edu}{lnguyen3@nd.edu}}
\affil{Joint Institute for Nuclear Astrophysics, Department of Physics, University of Notre Dame, Notre Dame, IN 46556}

\author{Nguyen-Anh Vinh}
\affil{Department of Physics, Hanoi National University of Education, 136 Xuan Thuy, Cau Giay, 1000, Hanoi, Vietnam}

\author{Yun-Jeong Kim}
\affil{Chungnam National University, Daejeon, 34113, South Korea}

\begin{abstract}
Electric dipole emission from rapidly spinning polycyclic aromatic hydrocarbons (PAHs) is widely believed as an origin of anomalous microwave emission (AME), but recently it encounters a setback due to the non-correlation of AME with PAH abundance seen in a full-sky analysis. Microwave observations for specific regions with well-constrained PAH features would be crucial to test the spinning dust hypothesis. In this paper, we present physical modeling of microwave emission from spinning PAHs from protoplanetary disks (PPDs) around Herbig Ae/Be stars and T-Tauri stars where PAH features are well observed. Guided by the presence of 10 $\mum$ silicate features in some PPDs, we also model microwave emission from spinning nanosilicates. Thermal emission from big dust grains is computed using the Monte Carlo radiative transfer code (\textsc{radmc-3d}; \citealt{2012ascl.soft02015D}). Our numerical results demonstrate that microwave emission from either spinning PAHs or spinning nanosilicates dominates over thermal dust at frequencies $\nu< 60$ GHz, even in the presence of significant grain growth. Finally, we attempt to fit mm-cm observational data with both thermal dust and spinning dust for several disks around Herbig Ae/Be stars that exhibit PAH features and find that spinning dust can successfully reproduce the observed excess microwave emission (EME). Future radio observations with ngVLA, SKA and ALMA Band 1 would be valuable for elucidating the origin of EME and potentially open a new window for probing nanoparticles in circumstellar disks. 
\end{abstract}

\section{Introduction\label{sec:intro}}

Polycyclic aromatic hydrocarbons (PAHs) are an important dust component of the interstellar medium (ISM, see review by \citealt{2008ARA&A..46..289T}). Following the absorption of ultraviolet (UV) photons, PAH molecules reemit radiation in mid-infrared, producing prominent 3.3, 6.2, 7.7, 8.6, 11.3, and 17 $\mu$m features  (\citealt{1984A&A...137L...5L}; \citealt{1985ApJ...290L..25A}; \citealt{2007ApJ...656..770S}; \citealt{2007ApJ...657..810D}). Rapidly spinning PAHs also emit electric dipole radiation in microwaves via a new mechanism, so-called spinning dust (\citealt{1998ApJ...508..157D}; \citealt{Hoang:2010jy}). The latter is the most likely origin of anomalous microwave emission (AME) that contaminates Cosmic Microwave Background (CMB) radiation  (\citealt{Kogut:1996p5293}; \citealt{Leitch:1997p7359}; \citealt{PlanckCollaboration:2011hw}; \citealt{2016A&A...594A..10P}). 

PAH molecules appear to be a natural carrier of the AME (\citealt{1998ApJ...508..157D}; \citealt{PlanckCollaboration:2011hw}) because it is an established component of interstellar dust \citep{2007ApJ...663..866D}. Yet such an explanation recently faces a setback due to no correlation between the observed AME and PAH abundance based on a full-sky analysis by \cite{2016ApJ...827...45H}.  Due to the spatial variation of PAH properties (e.g., geometry, size, and electric dipole moment), it is rather challenging to achieve a robust constraint for the carrier of AME by means of the full-sky analysis (see \citealt{Dickinson:2018ix} for a review). Therefore, observations of AME from specific regions with well-constrained PAH properties are critical to elucidate the exact carrier of AME. 

PAH molecules are usually detected in circumstellar disks around Herbig Ae/Be stars and some T-Tauri stars (\citealt{2004A&A...427..179H}; \citealt{2017ApJ...835..291S}). Their presence in PPDs is puzzling because one expects that interstellar PAHs are already depleted in dense cores due to coagulation. Thus, PAH molecules in PPDs may be newly formed particles as a result of dynamical processes, such as desorption of PAHs from the grain surface due to stellar radiation heating, replenishment due to collisions of planetesimals. PAHs can be formed in PPDs near the high temperature and density regions (see \citealt{2011EAS....46..271K}). Interestingly, if PAHs are produced by fragmentation of big carbonaceous grains, then, we expect a population of silicate nanoparticles that can also be produced by fragmentation of big silicate grains. Note that the modeling of very small grains (hereafter VSGs) around Herbig Ae/Be stars is previously studied in \cite{1993A&A...275..527N}. Alternatively, nanoparticles (including PAHs and nanosilicates) may follow a different evolution from classical grains (size of $0.1\mu$m). Thus, while classical dust grains are depleted in the disk due to coagulation and settling, PAHs/VSGs that are well mixed to the gas can exempt from grain settling and coagulation, and turbulence can be responsible for the mixing (see \citealt{2007A&A...473..457D}).

Modern understanding of AME indicates that, in addition to spinning PAHs, rapidly spinning silicate nanoparticles can successfully reproduce the observed AME in the diffuse ISM (\citealt{2016ApJ...824...18H}; \citealt{2017ApJ...836..179H}). Spinning iron nanoparticles cannot reproduce the entirety of the observed AME (\citealt{2016ApJ...824...18H}). Although, the presence of nanosilicates in the ISM remains a hypothesis, in contrast to PAHs, an analysis by \cite{Li:2001p4761} shows that the fraction of total Si abundance (Si/H$= 3.6\times 10^{-5}$) contained in ultrasmall grains, denoted by $Y_{\rm Si}$, can reach $Y_{\rm Si}\sim 10\%$ without violating the observational constraints of the UV starlight extinction and mid-infrared (IR) emission. \citealt{2016ApJ...824...18H} found that their emission and UV absorption do not violate the observational constraints for $Y_{\rm Si}< 15\%$. As PAHs, we expect nanosilicates are present in PPDs as a component of dust evolution model (see \citealt{2007prpl.conf..767N}). Indeed, \cite{2017ApJ...835..291S} found strong $9.7\mum$ emission Si-O features by nanosilicates in 40 out of 61 circumstellar disks (cf. \citealt{2008ApJ...684..411K}). 

\cite{2006ApJ...646..288R} carried out a simple modeling of microwave emission by spinning PAHs for the fiducial disks around T-Tauri ($M_{\star}<2M_{\odot}$) and Herbig A/Be stars ($2M_{\odot}\le M_{\star}<10M_{\odot}$). Assuming a one-dimensional disk structure, the author found that spinning dust emission dominates over thermal dust emission for $\nu<60$ GHz.

Radio observations of circumstellar disks often show excess emission at microwave frequencies, i.e., $\nu < 100$ GHz, above what is extrapolated from thermal dust emission at sub(mm) wavelengths, which we term excess microwave emission (hereafter EME).\footnote{EME is different from AME in the sense that the latter is the excess emission left after removing all three known galactic emission components, including thermal emission, free-emission, and synchrotron.} For instance, Very Large Array (VLA) observations by \cite{2002ApJ...568.1008C} and \cite{2004A&A...416..179N} reveal EME at 7 mm (or $\nu \sim 43$ GHz) from the disk around T-Tauri star, TW Hya, whereas \cite{2005ApJ...626L.109W} found excess emission at 3.5 cm (or 9 GHz). The authors explained EME by thermal emission from very big grains (i.e., cm-sized grains). EME is also detected in circumstellar disks around Herbig Ae/Be stars (\citealt{1993ApJS...87..217S}; \citealt{2006MNRAS.365.1283D}; \citealt{2011ApJ...727...26S}).  Although thermal dust from very big grains as well as free-free emission from winds are believed to be responsible for such excess emission, the exact mechanism is still unclear. Very recently, \cite{2017MNRAS.466.4083U} found EME in 11 disks around T-Tauri stars and suggested that multiple mechanisms different from thermal dust may be responsible for EME. To better understand the origin of EME, we will explore whether spinning dust could reproduce the observed EME.

With high resolution and low frequencies, next-generation VLA (ngVLA), ALMA Band 1, and SKA would be useful for observing spinning dust emission from circumstellar disks around T-Tauri and Herbig A/Be stars (\citealt{2013arXiv1310.1604D}; \citealt{Scaife:2013fu}). Radio observations by SKA and ALMA Band 1 would be crucial to study grain growth from mm to cm-sized pebbles as a first step of planet formation \citep{2015aska.confE.117T}. 

To provide a more realistic predictions for future observations, in this paper, we will improve modeling of microwave emission from \cite{2006ApJ...646..288R} by (1) treating the realistic geometry (i.e.,  two-dimensional) of disks, (2) finding the dust grain temperature using the publicly available 3D Monte Carlo radiative transfer code (\textsc{radmc-3d}; \citealt{2012ascl.soft02015D}),\footnote{The code and user guide are available at http://www.ita.uni-heidelberg.de/~dullemond/software/radmc-3d/.} (3) considering both emission from disk interior and surface layers, and (4) accounting for microwave emission from spinning nanosilicates. We also perform modeling of thermal dust emission with grain growth to 10 cm, in order to quantify the simultaneous effect of grain growth and spinning dust on the spectral energy density (SED).

The structure of this paper is as follows. Section \ref{sec:disk} describes the physical model of circumstellar disks. In Section \ref{sec:model}, we review the spinning dust model, and calculate the excitation coefficients for the disk conditions, which demonstrate that the collisions dominate both damping and excitation. Section \ref{sec:therm} presents dust opacity calculated for the different grain size distribution and thermal emission. Section \ref{sec:result} describes the SED from spinning dust for a wide range of disks. An extended discussion on implications of our results and especially an explanation of EME from circumstellar disks in terms of spinning dust are presented in Section \ref{sec:discuss}. A summary of our main results is presented in Section \ref{sec:sum}.

\section{Circumstellar Disk Model}\label{sec:disk}
\subsection{Gas density profile}
\begin{figure}
\includegraphics[width=0.48\textwidth]{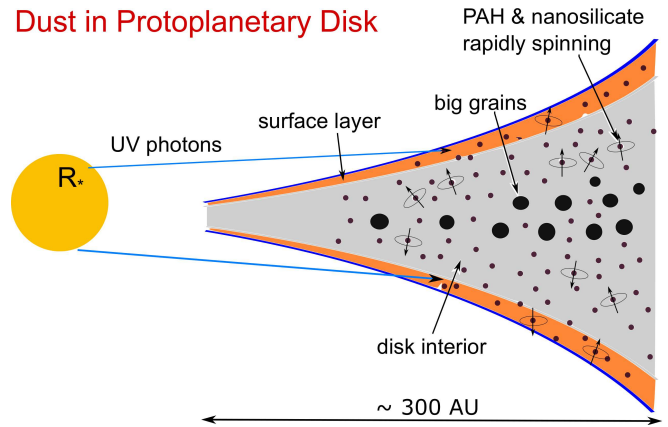}
\caption{Schematic illustration of a flared disk around a young star consisting of the surface layers and disk interior. The surface layers are directly heated by stellar radiation that can penetrate a thickness of optical depth $\tau_{V}=1$. The re-emission of the hot dust grains from the surface layers will heat big grains in the disk interior. Rapidly spinning PAHs/nanosilicates emit microwave radiation, while big grains emit primarily in far IR-(sub)mm.}
\label{fig:disk}
\end{figure}

This section briefly describes the disk model adopted for our modeling of spinning dust emission. We adopt a flared, radiative equilibrium disk model from \cite{1997ApJ...490..368C} (see also \citealt{2001ApJ...560..957D}). The schematic model of a protoplanetary disk is shown in Figure \ref{fig:disk}. PAHs and nanoparticles (i.e., VSGs) are assumed to be well mixed with the gas, thus present in the entire disk.

The total mass surface density at disk radius $R$ is given by
\bea
\Sigma(R)=\Sigma_{1}\left(\frac{R}{1~\AU}\right)^{-\alpha},\label{eq:sigmar}
\ena
where $\alpha$ is the model constant, and $\Sigma_{1}$ is the mass density at $R=1\AU$. When the total surface density at $R_{\rm out}$ is given as $\Sigma_{0}$, then, we have $\Sigma_{1}=\Sigma_{0}(R_{\rm out}/1\AU)^{-\alpha}$ (see Appendix \ref{apd:MPAH}). 

Assuming a Gaussian vertical profile, the gas density at radius $R$ for the hydrostatic disk model (\citealt{1974MNRAS.168..603L}) reads
\bea
n_{\H}(R,z)&\approx&\frac{1}{2m_{\H}}\frac{\Sigma(R)}{H_{p}\sqrt{2\pi}}\exp\left(-\frac{z^{2}}{2H_{p}^{2}}\right), \label{eq:ngas0}
\ena
where  the pressure height scale $H_{p}$ is described by
\bea
\frac{H_{p}}{R}=\frac{H_{0}}{R_{0}} \left(\frac{R}{R_{0}}\right)^{1./7},\label{eq:Hp}
\ena
{where $H_{0}$ is the aspect ratio at the reference radius $R_{0}$. For $R_{0} =100$ AU, $H_{0}/R_{0}$ is taken to be 0.1 as a fiducial model, which corresponds to $H_{p}/R=0.1\times 3^{1/7}$ at $R_{\rm out}=300$ AU. Although the chosen aspect ratio is much lower than predicted by \cite{1997ApJ...490..368C}, it is comparable to observations (\citealt{Avenhaus:2018tt}).}

The typical value $\alpha=1$ is adopted. Other physical parameters, including $R_{\rm in},R_{\rm out}$, are listed in Table \ref{tab:diskmod}.

\subsection{Gas and Dust temperatures}
Following the popular model of protoplanetary disks (\citealt{1997ApJ...490..368C}), the surface layer is defined by a path of optical depth $\tau_{V}=1$. At distance $r$ from the star, the surface layer is heated to a high temperature $T_{s}$ by stellar radiation. Subsequent collisions with gas atoms result in gas heating. Dust grains in these superheated layers reemit radiation in IR that in turn heats gas and dust in the disk interior to a temperature $T_{i}$. For an isothermal disk, gas and dust are in thermal equilibrium, so that $T_{gi}=T_{di}$. 

In our paper, instead of using the simplified temperature profile as in \cite{2006ApJ...646..288R}, we directly compute the dust temperature for the realistic disk model using \textsc{radmc-3d}. The dust opacity is calculated assuming a power grain size, with different values of $a_{\max}$ for silicate grains. 

\begin{figure*}
\includegraphics[width=0.48\textwidth]{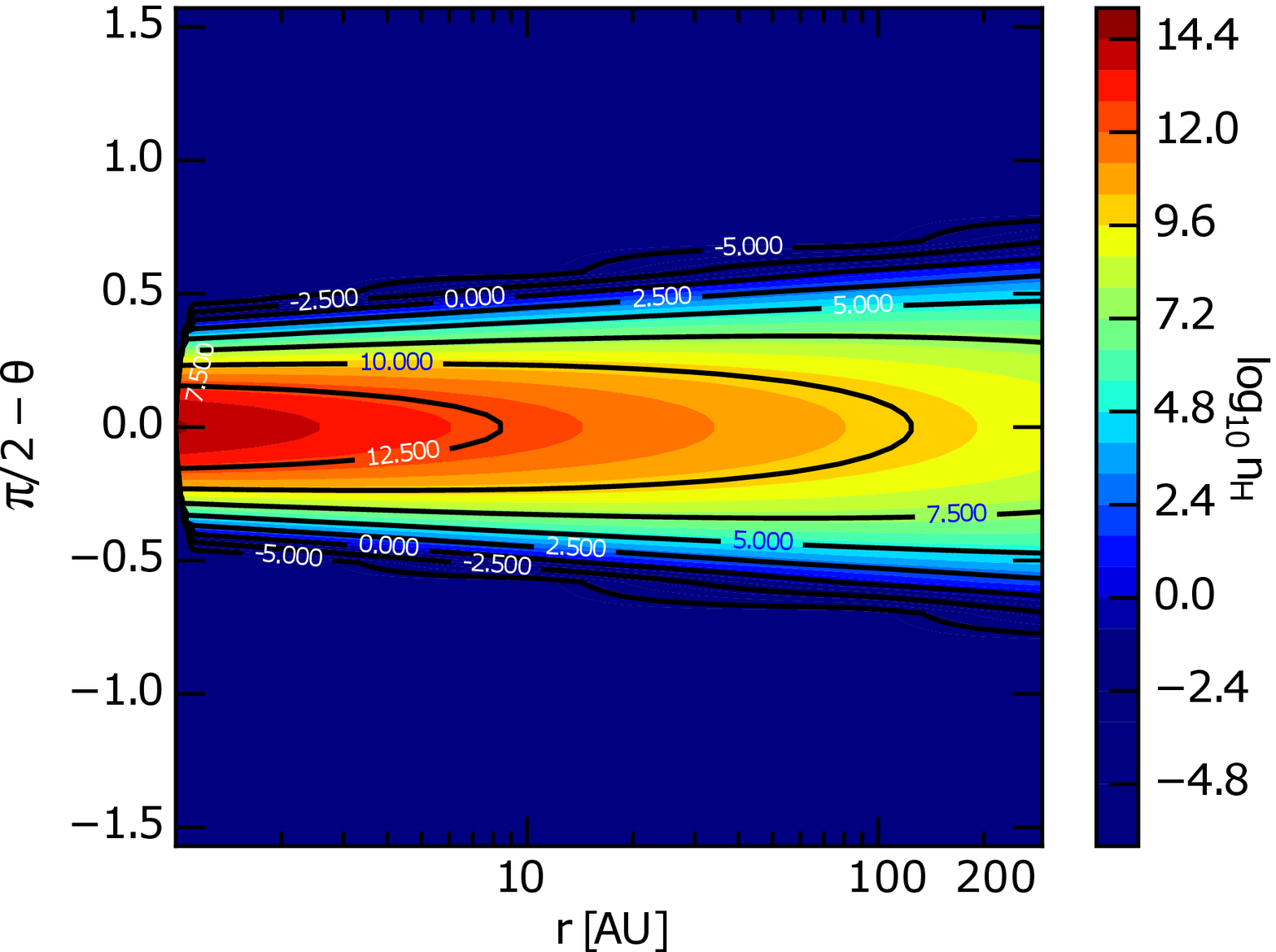}
\includegraphics[width=0.48\textwidth]{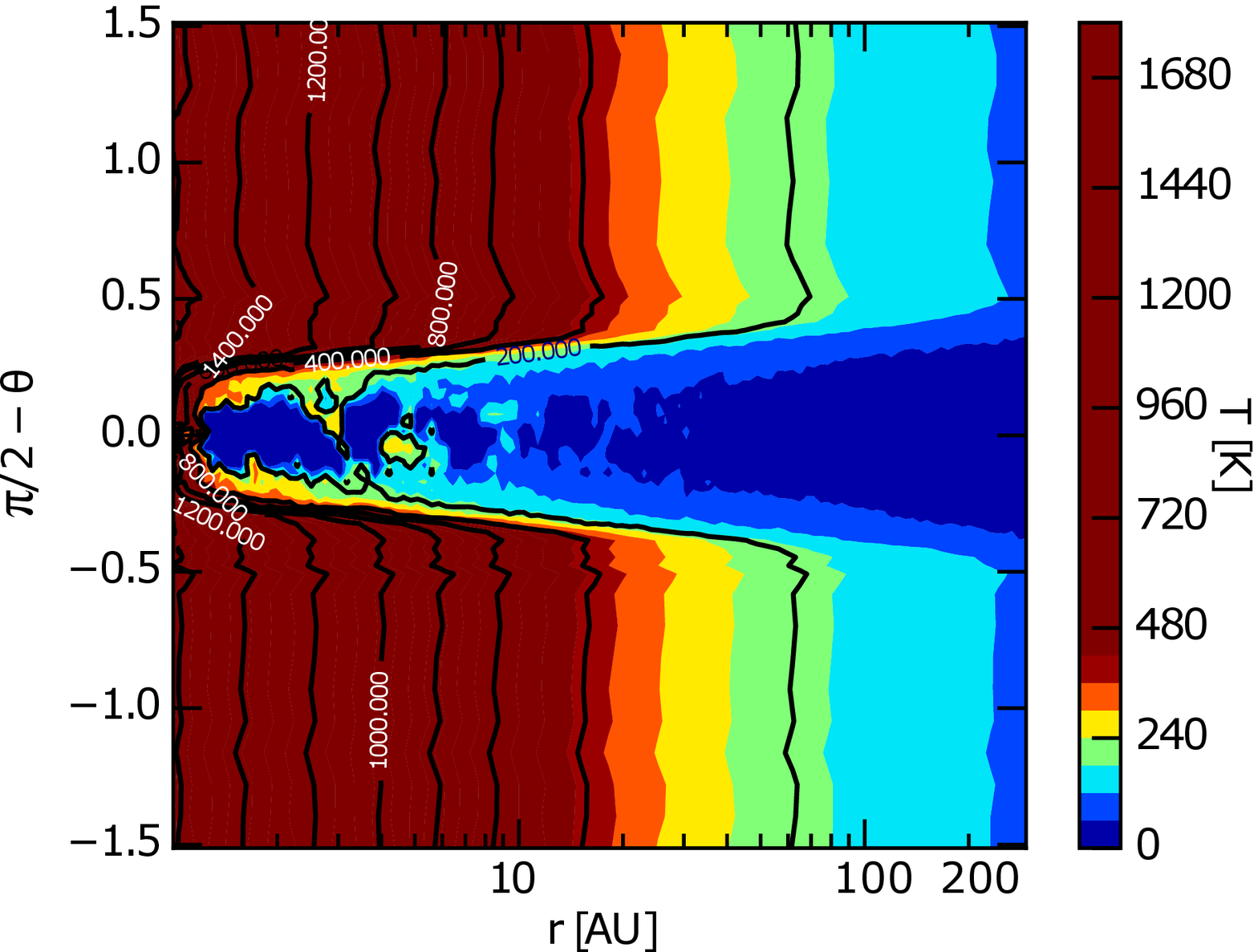}
\caption{Gas number density and dust temperature obtained from \textsc{radmc-3d} for a fiducial disk around Herbig AeBe stars, with $\Sigma_{1}=10^{3}\g\cm^{-2}$ and $\alpha=1$.}
\label{fig:nTgas}
\end{figure*}

Figure \ref{fig:nTgas} shows the gas density and temperature for a fiducial disk around Herbig Ae/Be stars. {The obtained dust temperature depends on $a_{\max}$ because the opacity $\kappa$.}

\subsection{Gas ionization and charge of PAHs/VSGs}
Gas in PPDs can be ionized by X-rays, far-UV photons, cosmic rays (see \citealt{2011ApJ...735....8P} and ref therein). Theoretical estimates provide the hydrogen ionization fraction $x_{\H}\sim 10^{-6}-10^{-5}$ for the surface layer and $x_{\H}\sim 10^{-10}$ for the disk interior \citep{2011ApJ...735....8P}.

The ionization of PAHs/VSGs can be approximately described by a three-layer vertical structure model (\citealt{2007A&A...466..229V}). In the surface layers, PAHs/VSGs are positively charged due to photoelectric emission induced by stellar UV photons. In the intermediate region, PAHs/VSGs are mostly neutral, reflecting the balance of photoelectric emission and electron captures, and PAHs/VSGs are negatively charged in the diskplane due to electron collisions and the lack of UV photons (see \citealt{2011EAS....46..271K}; also \citealt{2014A&A...563A..78M}). Nevertheless, as shown in the next section, the effect of PAH and nanoparticles charge states is not important for grain rotation in very dense environments as PPDs.

\section{Spinning dust model}\label{sec:model}
\subsection{Electric dipole moment and emission power}
The rotational emission mechanism is built upon the assumption that nanoparticles own non-zero electric dipole moments. PAH molecules can acquire intrinsic dipole moments due to polar bonds (see \citealt{1998ApJ...508..157D}). The attachment of SiO and SiC molecules to the grain surface gives rise to the electric dipole moment for nanosilicates (\citealt{2016ApJ...824...18H}). 

Let $N$ be the total number of atoms in a nanoparticle of effective size $a$ that is defined as the radius of an equivalent sphere of the same volume. Assuming PAHs with a typical structure C:H=$3:1$ having mean mass per atom $m\approx 9.25$ amu, one obtains $N=545a_{-7}^{3}$ with $a_{-7}=a/10^{-7}\cm$, for the mass density $\rho=2\g\cm^{-3}$ (\citealt{1998ApJ...508..157D}). Assuming nanosilicate with a structure SiO$_{4}$Mg$_{1.1}$Fe$_{0.9}$ having $m=24.15$ amu, one has $N=418a_{-7}^{3}$ for $\rho=4\g\cm^{-3}$ \citep{2016ApJ...824...18H}.

Let $\beta$ be the dipole moment per atom in the grain. Assuming that dipoles have a random orientation distribution, the intrinsic dipole moment of the grain can be estimated using the random walk formula:
\bea
\mu^{2}=N\beta^{2}\simeq 86.5(\beta/0.4\D)^{2} a_{-7}^{3} \D^{2}\label{eq:muin}
\ena
for PAHs, and $\mu^{2}\simeq 66.8(\beta/0.4\D)^{2} a_{-7}^{3} \D^{2}$ for nanosilicates \citep{2016ApJ...824...18H}.

The power emitted by a rotating dipole moment $\mu$ at angular velocity $\omega$ is given by the Larmor formula:
\bea
P(\omega,\mu)=\frac{2}{3}\frac{\omega^4\mu^2\sin^2\theta}{c^3}~~~,
\ena
where $\theta$ is the angle between $\bomega$ and $\bmu$. Assuming an uniform distribution of the dipole orientation, $\theta$, then, $\sin^{2}\theta$ is replaced by $\langle \sin^{2}\theta\rangle=2/3$.

\subsection{Rotational damping and excitation coefficients}
Rotational damping and excitation for nanoparticles, in general, arise from collisions between the grain and gaseous atoms (neutrals and ions) followed by the evaporation of atoms/molecules from the grain surface, absorption of starlight and IR emission (\citealt{1998ApJ...508..157D}; \citealt{Hoang:2010jy}). Moreover, the distant interaction between the grain electric dipole and electric field induced by passing ions results in an additional effect, namely plasma drag. The rotational damping and excitation for these processes are respectively described by the dimensionless damping coefficient $F_{j}$ and $G_{j}$ where $j=n,i,p,IR$ denotes the neutral-grain collision, ion-grain collision, plasma drag, and IR emission (see \citealt{Hoang:2010jy}). 

We consider the major neutral components in the PPDs, including H, H$_{2}$, He, and ions H$^{+}$ and  C$^{+}$. The typical values $x_{\H}=10^{-8}, n(\rm He)\sim 0.1n_{\H}, y=2n(\H_{2})/n_{\H}=1, x_{M}=n(C^{+})/n_{\H}=10^{-8}$.

Let $T_{\rm rot}$ be the rotational temperature of spinning nanoparticles, so that $3kT_{\rm rot} = I_{1}\langle \omega^{2}\rangle$. Thus, using the rms angular velocity from \cite{1998ApJ...508..157D}, we obtain
\bea
\frac{T_{\rm rot}}{T_{\rm gas}}=\frac{G}{F}\frac{2}{1 + [1+ (G/F^2)(20\tH/3\ted)]^{1/2}},\label{eq:Trot}
\ena
where $\tH$ and $\ted$ are the characteristic damping times due to gas collisions and electric dipole emission (see \citealt{1998ApJ...508..157D}). From Figure \ref{fig:disk} we see that the majority of the disk has $n_{\H}>10^{5}\cm^{-3}$, which results in $\tau_{\ed}/\tau_{\H}\sim (a/3.5\AA)^{7}(n_{\H}/10^{4}\cm^{-3}) \gg 1$ (see \citealt{Hoang:2010jy}). Thus, $T_{\rm rot}/T_{\rm gas}\sim G/F$, i.e., the rotational temperature is only determined by $F$ and $G$ coefficients.

\begin{figure*}
\centering
\includegraphics[width=0.9\textwidth]{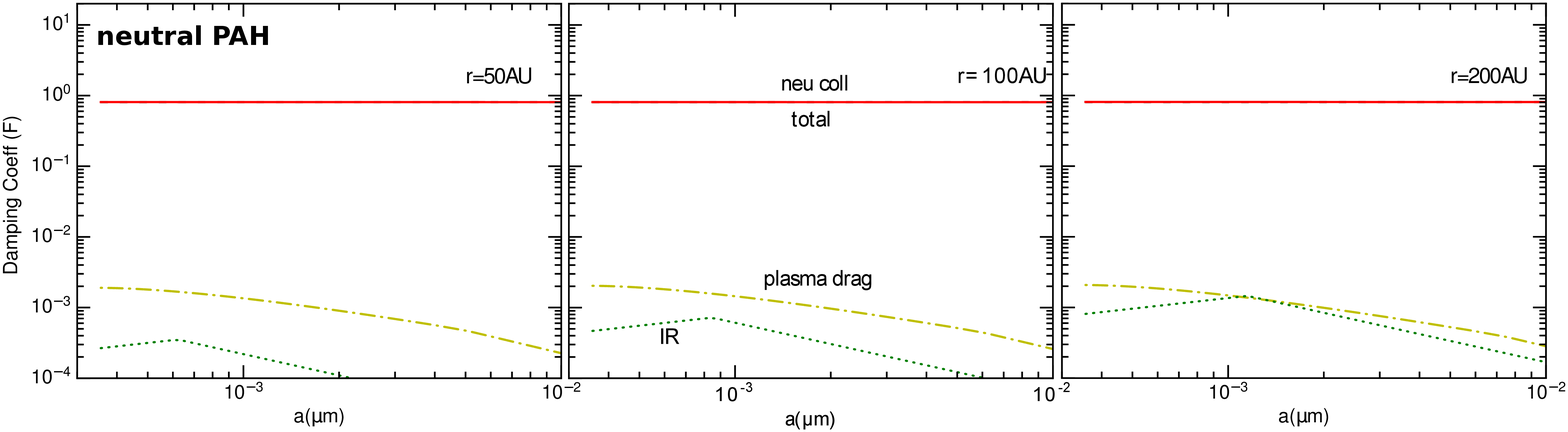}
\includegraphics[width=0.9\textwidth]{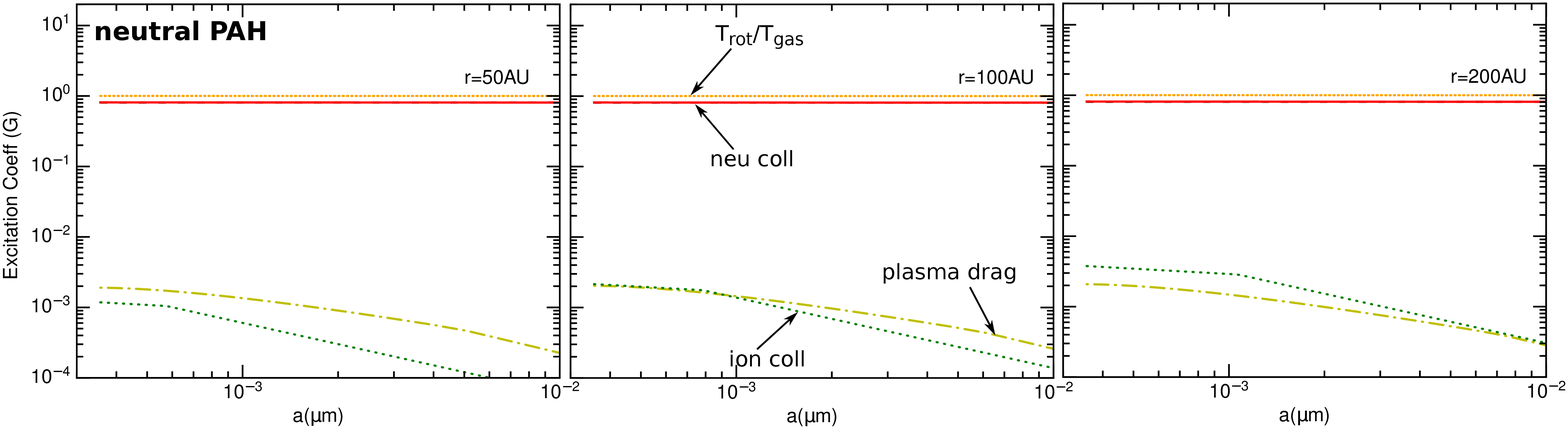}
\caption{Damping and excitation coefficients for neutral PAHs ($Z=0$) at $r=50$ AU (left panels), $100$ AU (middle) and $r= 200$ AU (right panels). The ionization $x_{\H}=10^{-8}$ is adopted. Neutral collisions dominate both damping and excitation. The rotation is thermal, with $T_{\rm rot}/T_{\rm gas}=1$.}
\label{fig:FG_coeff}
\end{figure*}

\begin{figure*}
\centering
\includegraphics[width=0.9\textwidth]{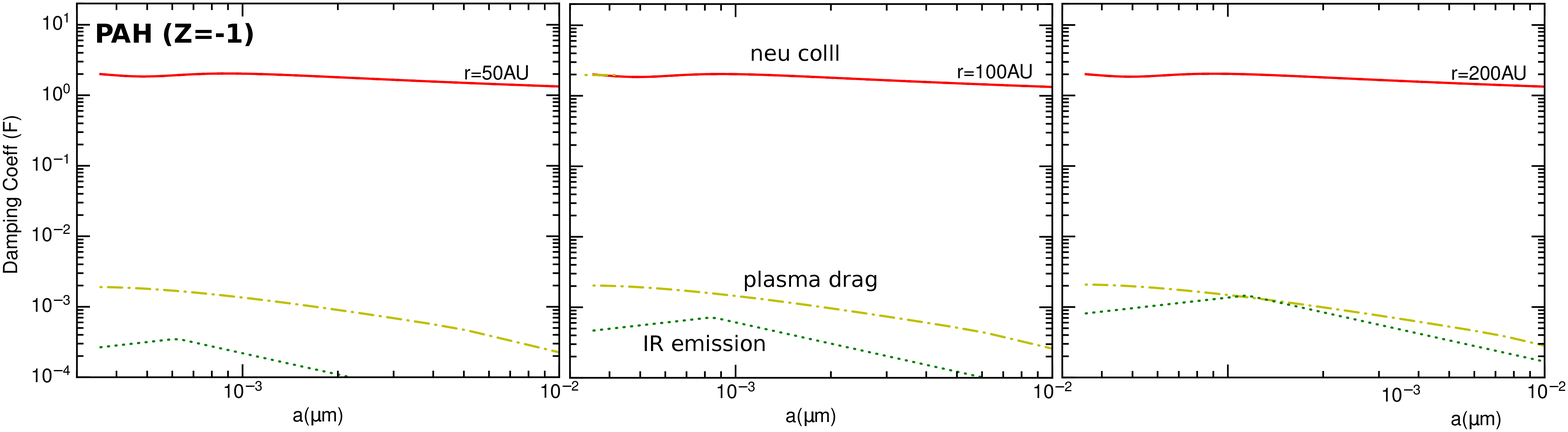}
\includegraphics[width=0.9\textwidth]{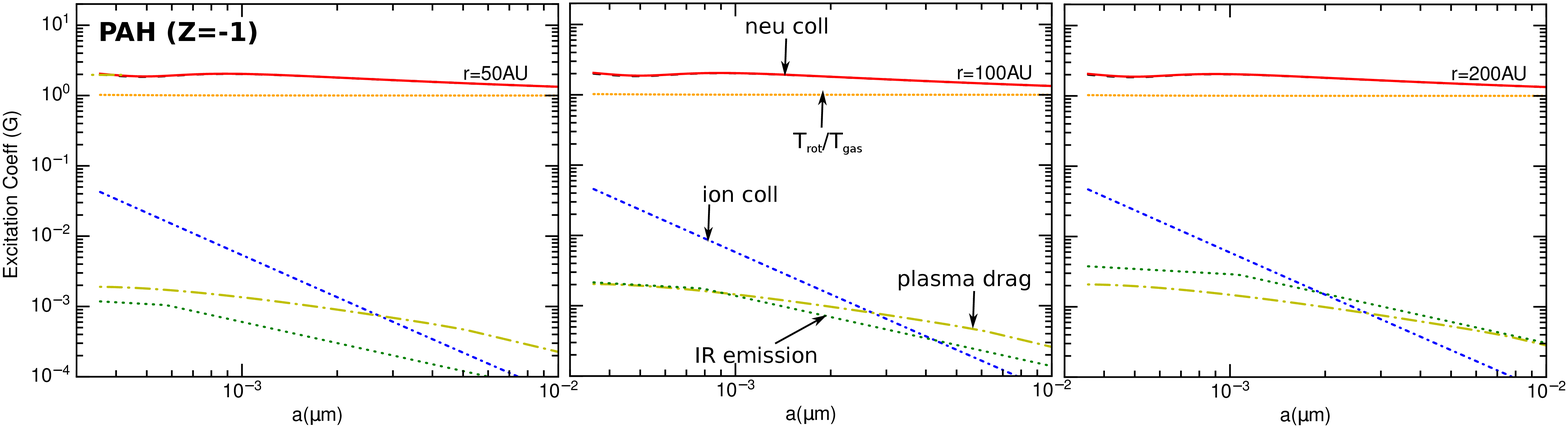}
\caption{Same as Figure \ref{fig:FG_coeff}, but for negatively charged PAHs ($Z=-1$). Neutral-grain collision is enhanced due to dipole interaction, but the rotation of small PAHs is thermal with $T_{\rm rot}/T_{\rm gas}=1$.}
\label{fig:FG_coeff1}
\end{figure*}

Figure \ref{fig:FG_coeff} shows the $F$ and $G$ coefficients for neutral PAHs at three locations in the diskplane at $50, 100~\AU$ and $200~\AU$. The corresponding gas density is $n_{\H}=10^{8}, 10^{9}$ and $10^{10}\cm^{-3}$ (see Equations \ref{eq:ngas0}). The radiation intensity factor is $U=4\times 10^{4}, 2\times 10^{4}$ and $10^{3}$ (see Eq. \ref{eq:U}). A typical ionization fraction $x_{\H}=10^{-8}$ is chosen. In all three locations, collisional interactions with neutral dominate the damping and excitation. In Figure \ref{fig:FG_coeff1}, we show the results for negative charged PAHs. Collisions still dominate the interaction, such that the evaporation is $T_{\rm ev}=T_{\rm gas}$, leading to the detailed balance with $F_{n}=G_{n}$.

In the disk interior with anion PAHs, the ionization fraction is too low, i.e, $x_{\H}<10^{-8}$, ion excitation and plasma drag are not important, as shown in Figure \ref{fig:FG_coeff}. In the surface layer with higher ionization fraction ($x_{\H}\sim 10^{-5}$), rotational excitation by neutral-positively charged PAH can be efficient. Nevertheless, the mass of the surface layer is very low compared to the total disk mass, such that the ionization effect has little impact on the net spinning dust emission. As a result, in the following, we can adequately assume $T_{\rm rot}=T_{\gas}$ for modeling spinning dust throughout the disk.

\subsection{Angular momentum distribution function}
In high-density conditions where collisional excitations dominate rotation of nanoparticles (e.g., in PPDs), the grain angular momentum can be appropriately described by the Maxwellian distribution: 
\bea
f_{\omega}(\omega, T_{\rm rot})=\frac{4\pi}{ (2\pi)^{3/2}}\frac{I_{1}^{3/2}\omega^{2}}{(kT_{\rm rot})^{3/2}}\exp\left(-\frac{I_{1}\omega^{2}}{2kT_{\rm rot}} \right),\label{eq:fomega}
\ena
where $I_{1}=8\pi \rho a^{5}/15$ is the moment of inertia of the spherical nanoparticle of mass density $\rho$.

\subsection{Size distribution: PAHs and nanosilicates}

Following \cite{Li:2001p4761}, nanoparticles are assumed to follow a log-normal size distribution:
\bea
\frac{1}{n_{\H}}\frac{dn_{j}}{da} = \frac{B_{j}}{a}\exp\left(-0.5\left[\frac{\log (a/a_{0,j})}{\sigma_{j}}\right]^{2}\right) ,\label{eq:dnda_log}
\ena
where $j=PAH, sil$ corresponds to PAHs and nanosilicate composition, $a_{0,j}$ and $\sigma_{j}$ are the model parameters, and $B_{j}$ is a constant determined by
\bea
B_{j}&=&\frac{3}{(2\pi)^{3/2}}\frac{{\rm exp}(-4.5\sigma_{j}^{2})}{\rho \sigma a_{0,j}^{3}}\nonumber\\
&&\times \left(\frac{m_{X}b_{X}}{1+{\rm erf}[3\sigma/\sqrt{2} + {\rm ln}(a_{0}/a_{\min})/\sigma\sqrt{2}}\right),\label{eq:Bconst}
\ena
where $b_{X}=X_{\H}Y_{X}$ with $Y_{X}$ being the fraction of $X$ abundance contained in very small sizes and $X_H$ being the solar abundance of element $X$, and $m_{X}$ is the grain mass per X atom. In our studies, $X=$C for PAHs and $X=$Si for nanosilicates. 

The peak of the mass distribution $a^{3}dn_{j}/d\ln a$ occurs at $a_{p}=a_{0,j}e^{3\sigma_{j}^{2}}$. Three parameters determine the size distribution of nanoparticles, including $a_{0,j},\sigma_{j}, Y_{X}$.

In realistic environments, $b_{X}$ should depend on the local conditions and is a function of the radial distance $R$. However, due to poorly known nanoparticles in the disk, $b_{X}$ is kept constant in this paper.

\subsection{Spinning dust emissivity and emission spectrum}
Let $j_{\nu}^{a}(\mu, T_{\rm rot})$ be the emissivity from a spinning nanoparticle of size $a$ at location $(r,\theta,\phi)$ in the disk, where $T_{\rm rot}$ in general is a function of the local conditions. Thus,
\bea
j_{\nu}^{a}(\mu, T_{\rm rot})= \frac{1}{4\pi}P(\omega,\mu) 2\pi f_{\omega}(\omega,T_{\rm rot}),\label{eq:jem_a}
\ena
where $f_{\omega}$ is given by Equation (\ref{eq:fomega}).

The rotational emissivity per H nucleon is obtained by integrating over the grain size distribution (see \citealt{2011ApJ...741...87H}):
\bea
\frac{j_{\nu}(\mu, T_{\rm rot})}{n_{\H}}=\int_{a_{\min}}^{a_{\max}}j_{\nu}^{a}(\mu,T_{\rm rot})\frac{1}{n_{\H}} \frac{dn}{da} da,\label{eq:jem}
\ena 
where $dn/da = dn_{\rm PAH, sil}/da$ for spinning PAHs and nanosilicates, respectively.

Thus, the total emission luminosity from the disk is given by
\bea
L_{\nu, \rm sd}&=& \int_{R_{\rm in}}^{R_{\rm out}}r^{2} dr \int_{0}^{\pi}\sin\theta d\theta\int_{0}^{2\pi}d\phi n_{\H}(r,\theta,\phi)\frac{4\pi j_{\nu}(\mu,T_{\rm rot})}{n_{\H}},\nonumber\\
&=& \int_{R_{\rm in}}^{R_{\rm out}}2\pi R dR \int dz n_{\H}(R,z)\left(\frac{4\pi j_{\nu}(\mu,T_{\rm rot})}{n_{\H}}\right),\label{eq:Fsd}
\ena
where $n_{\H}(R,z)$ is given by Equation (\ref{eq:ngas0}). For a disk at distance $D$ from the observer, the spectral flux density of spinning dust $F_{\rm sd}=L_{\nu, \rm sd}/4\pi D^{2}$.

\section{Thermal dust emission}\label{sec:therm}
\subsection{Dust opacity}
Let $Q_{\rm abs}(a,\nu)$ be the absorption efficiency for a grain of radius $a$ at frequency $\nu$. The density of {\it dust grains} is given by the grain size distribution $dn_{\rm gr}/da$. The dust opacity, defined as the total absorption cross-section per unit of dust mass, is given by
\bea
\kappa_{\rm abs}(\nu) = \frac{\int_{a_{\rm min}}^{a_{\max}} \pi a^{2}Q_{\rm abs}(a,\nu)(dn_{\rm gr}/da) da}{\int_{a_{\rm min}}^{a_{\max}} (4\pi \rho a^{3}/3)(dn_{gr}/da) da},\label{eq:kappa}
\ena
where $a_{\min, \max}$ are the lower and upper cutoffs of the size distribution of big grains. Here $a_{\rm min}=0.01\mum$ is chosen, and $a_{\rm max}$ is varied to account for grain growth.

We compute the absorption cross-section for spherical grains using the Mie theory coded from \cite{1983asls.book.....B}, assuming the optical constant of amorphous silicate (Mg$_{0.7}$Fe$_{0.3}$SiO$_{3}$) \footnote{http://www.astro.uni-jena.de/Laboratory/OCDB/data/silicate/amorph/pyrmg70.lnk}. The opacity is then calculated by Equation (\ref{eq:kappa}), assuming a power law of the grain size distribution $dn_{\rm gr}/da \sim a^{-q}$. 

\begin{figure}
\includegraphics[width=0.48\textwidth]{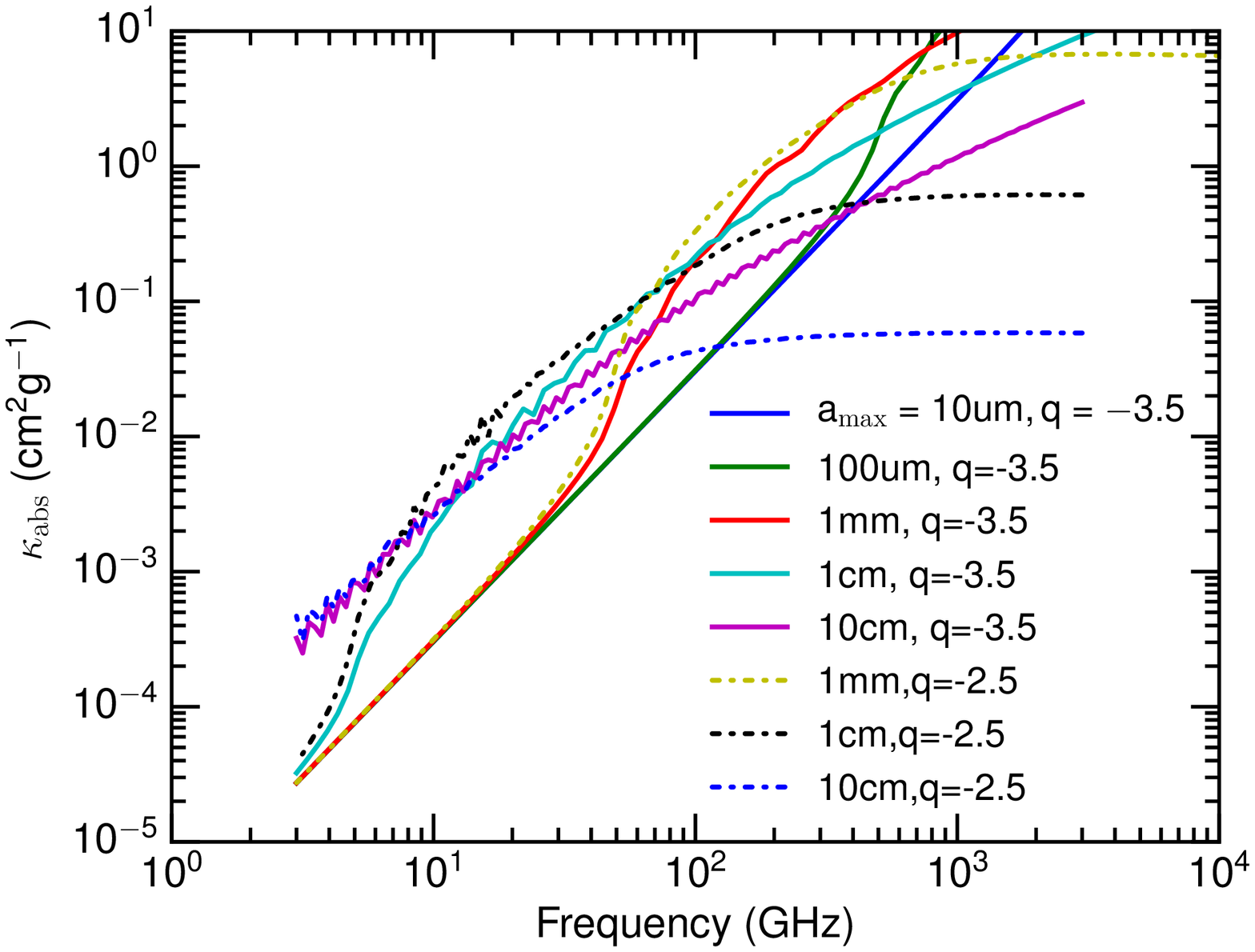}
\caption{Dust opacity of amorphous silicate computed for the different maximum grain sizes, assuming the power law size distribution $dn/da \propto a^{-q}$. The slope is more shallow for larger grains, but it does not follow the power law. }
\label{fig:kappa_abs}
\end{figure}

Figure \ref{fig:kappa_abs} shows the dust opacity for the different values of $a_{\max}$, assuming the typical value $q=-3.5$ and a more shallow distribution of $q=-2.5$. The grain growth from 1 cm to 10 cm can increase the opacity at $\nu < 30$ GHz by a factor of 2. The distribution of $q=-2.5$ results in the increase of the opacity at $\nu\sim 10-100$ GHz. Note that the opacity at $\nu=30$ GHz does not increase monotonically with $a_{\max}$.

\subsection{Thermal dust emission}
In the case of an isothermal disk along the vertical direction, the spectral flux density of thermal emission from the disk in the optically thin regime can be calculated by (see \citealt{2006ApJ...646..288R})
\bea
F_{\nu,\rm th}= \frac{1}{4\pi D^{2}}\int_{R_{\rm in}}^{R_{\rm out}}\pi B_{\nu}(T_{d})(1-e^{-\kappa_{\rm abs}\Sigma_{d}(R)})2\pi RdR,~~~~\label{eq:Fth}
\ena
where $\kappa_{\rm abs}(\nu)$ is given by Equation (\ref{eq:kappa}). 

In the present paper, we directly compute $F_{\nu,\rm th}$ using \textsc{radmc-3d} for the different grain size distributions and $a_{\max}$. This allows us to relate the SED of thermal dust to the effect of grain growth.

\section{Spinning dust emission spectrum from circumstellar disks}\label{sec:result}

\subsection{Numerical method and Model Setup}
\begin{figure}
\includegraphics[width=0.5\textwidth]{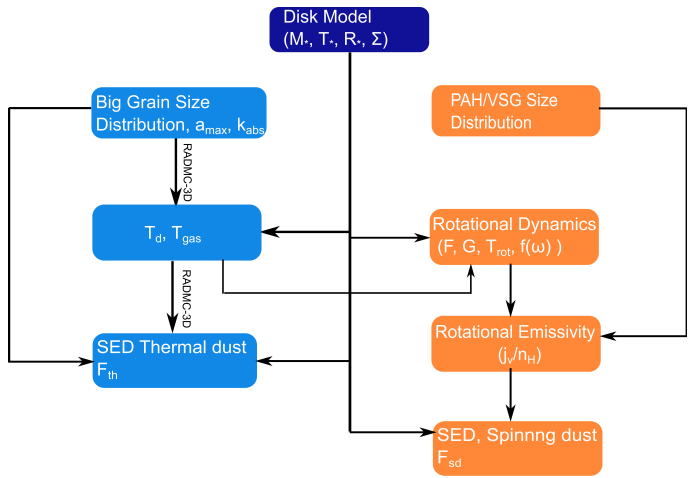}
\caption{Working diagram for modeling spinning dust and thermal dust emission from circumstellar disks.}
\label{fig:model}
\end{figure}

Our modeling strategy is depicted in Figure \ref{fig:model}. We adopt the fiducial model of a circumstellar disk around a Herbig Ae/Be star and T-Tauri star, with physical parameters listed in Table \ref{tab:diskmod}. For a set of the disk parameters, $T_{\star}, R_{\star}, \alpha$, we create a disk physical model as described in Section \ref{sec:disk} to generate the gas density profile $n_{\H}(r,\theta,\phi)$. We then use \textsc{radmc-3d} to calculate $T_{d}(r,\theta,\phi)$ for the constructed disk. {We consider the different grain size distributions, which have opacity given by Figure \ref{fig:kappa_abs}. Viscous heating and internal heating are not considered. For MC simulations, we use the default value of $N_{\rm phot}\sim 10^{5}$ photon packages.}
The grid resolutions are $N_{r}=128,~ N_{\theta}=32,~ n_{\phi}=128$, in which $R$ spans $R_{\rm in}$ to $R_{\rm out}$, $\theta$ from $\pi/3$ to $2\pi/3$, and $\phi$ from $0$ to $2\pi$. 

\begin{table}
\caption{Fiducial disk models}\label{tab:diskmod}
\begin{tabular}{llllllll} 
\hline\hline
{\it Objects} & $a_{0}$ & $\sigma$ & $T_{\star}$ & $M_{\star}$ & $R_{\star}$ & $R_{\rm in}$ & $R_{\rm out}$ \\
 & $(\AA)$ &  & $(\K)$ & $(M_{\odot})$ & $(R_{\odot})$ & $(\AU)$ & $(\AU)$ \\[1mm]

\hline\\
Herbig Ae/Be & 3.0	&  0.3 & 10000	& 2 & 2  &  1 & 300\\[1mm]
T-Tauri	 & 3.0 & 0.3 & 4000	&  0.5 & 2 &  0.1 & 300 \\[1mm]
\hline\hline\\
\end{tabular}
\end{table}

At each location $(r, \theta,\phi)$ with given local physical parameters ($n_{\H},T_{\rm gas}$), we can calculate the damping and excitation coefficients $F$ and $G$ to obtain $T_{\rm rot}$ using Equation (\ref{eq:Trot}). This process can be simplified by the fact that, in the dense conditions,  $T_{\rm rot}\sim T_{\gas}$. The spinning dust emissivity $j_{\nu}(r,\theta,\phi)/n_{\H}$ is then calculated using Equation (\ref{eq:jem}). Finally, the energy flux density of spinning dust is calculated by integrating over the symmetric disk as given by Equation (\ref{eq:Fsd}).

\subsection{Microwave emission from spinning PAHs}
We first consider the emission from spinning PAHs. The PAH size distribution is varied from $(a_{0}, \sigma)=(0.2\AA, 0.2)$ to $(0.5\AA,  0.5)$. Here we fix the C abundance contained in PAHs, $f_{\rm C}$, to be similar to the diffuse ISM, of $f_{\rm C}\sim 0.05$ (see \citealt{2007ApJ...657..810D}).\footnote{The effect of varying $f_{\rm C}$ is analogous to spinning nanosilicates, which will be quantified in the next section.} The lower and upper cutoff of the PAH size distribution $a_{\min}=3.5$ \AA~ and $a_{\max}=100$ \AA.

\begin{figure*}
\includegraphics[width=0.5\textwidth]{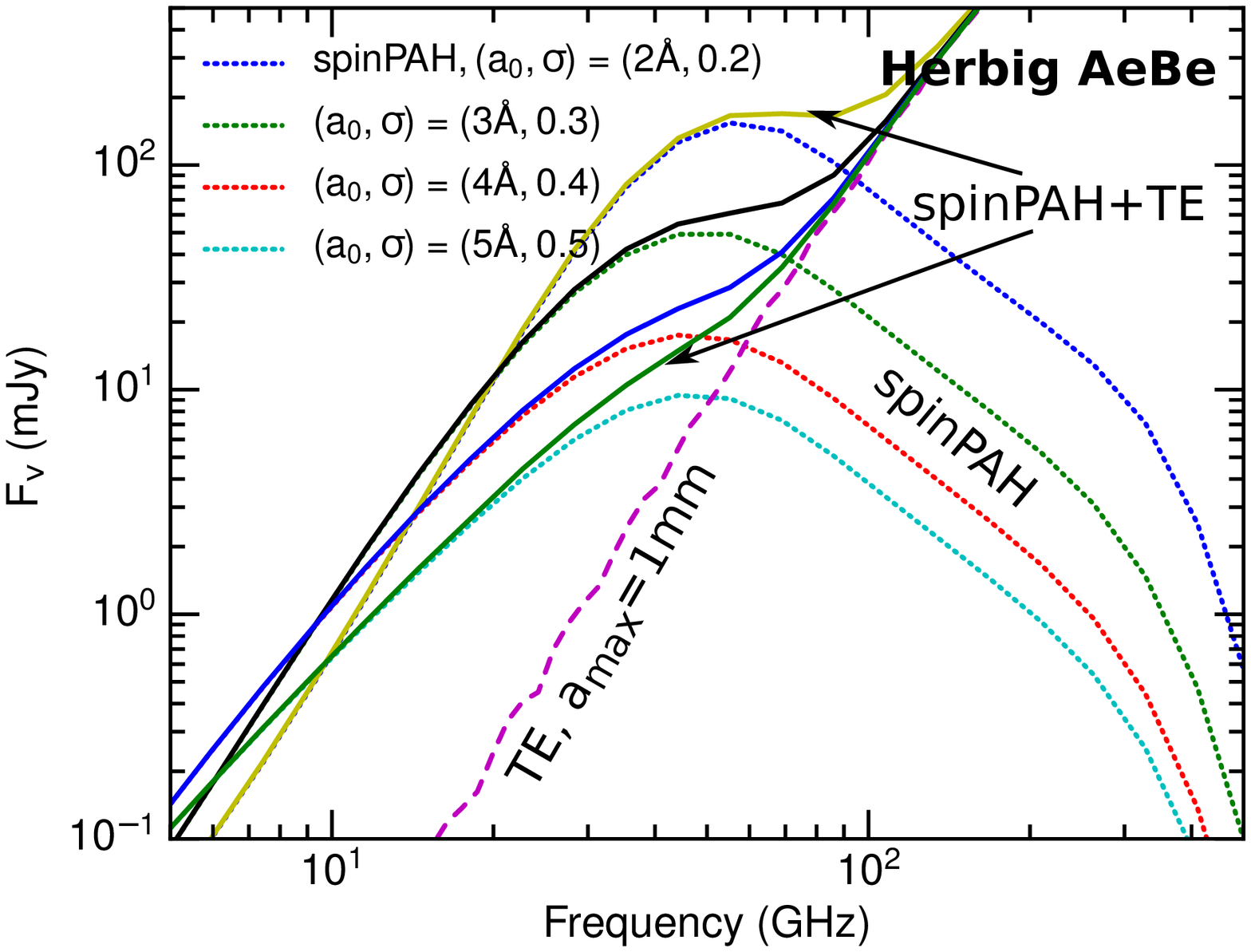}
\includegraphics[width=0.5\textwidth]{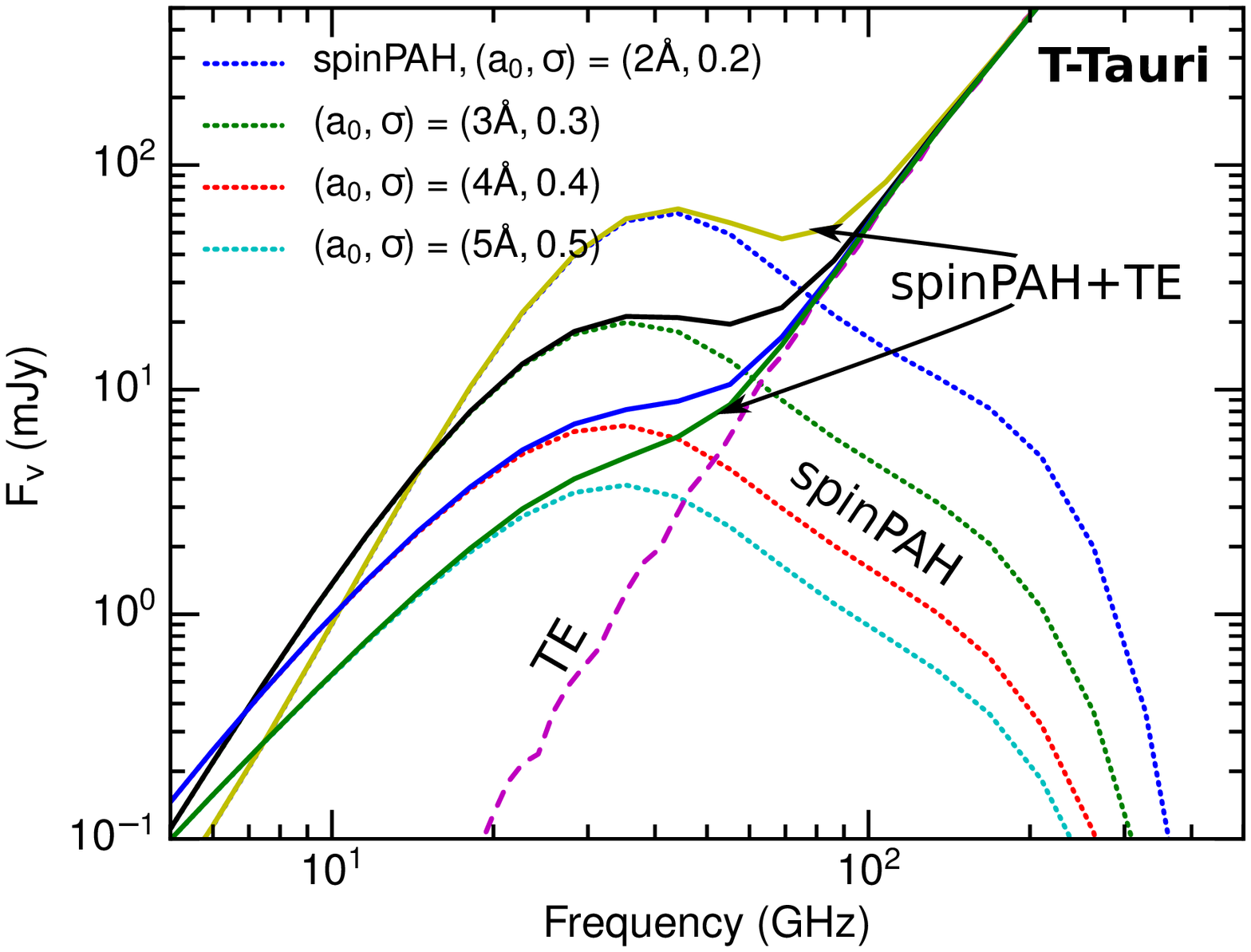}
\caption{Spectral flux density from spinning PAHs (dotted lines) of the different size distribution parameters ($a_{0},\sigma$), thermal dust emission for $a_{\max}=1$ mm (dashed line), and total emission from both spinning dust and thermal dust (solid lines). Left panel and right panel show the results for Herbig Ae/Be disk and T-Tauri disk, respectively.}
\label{fig:Fspd_disk_PAH}
\end{figure*}

Figure \ref{fig:Fspd_disk_PAH} shows the spectral flux density of spinning PAH emission from both the disk interior and surface layer for a Herbig Ae/Be (upper panels) and T-Tauri (lower panels) disks. Models with smaller PAHs tend to have stronger emission and higher peak frequency.

\begin{figure*}
\includegraphics[width=0.5\textwidth]{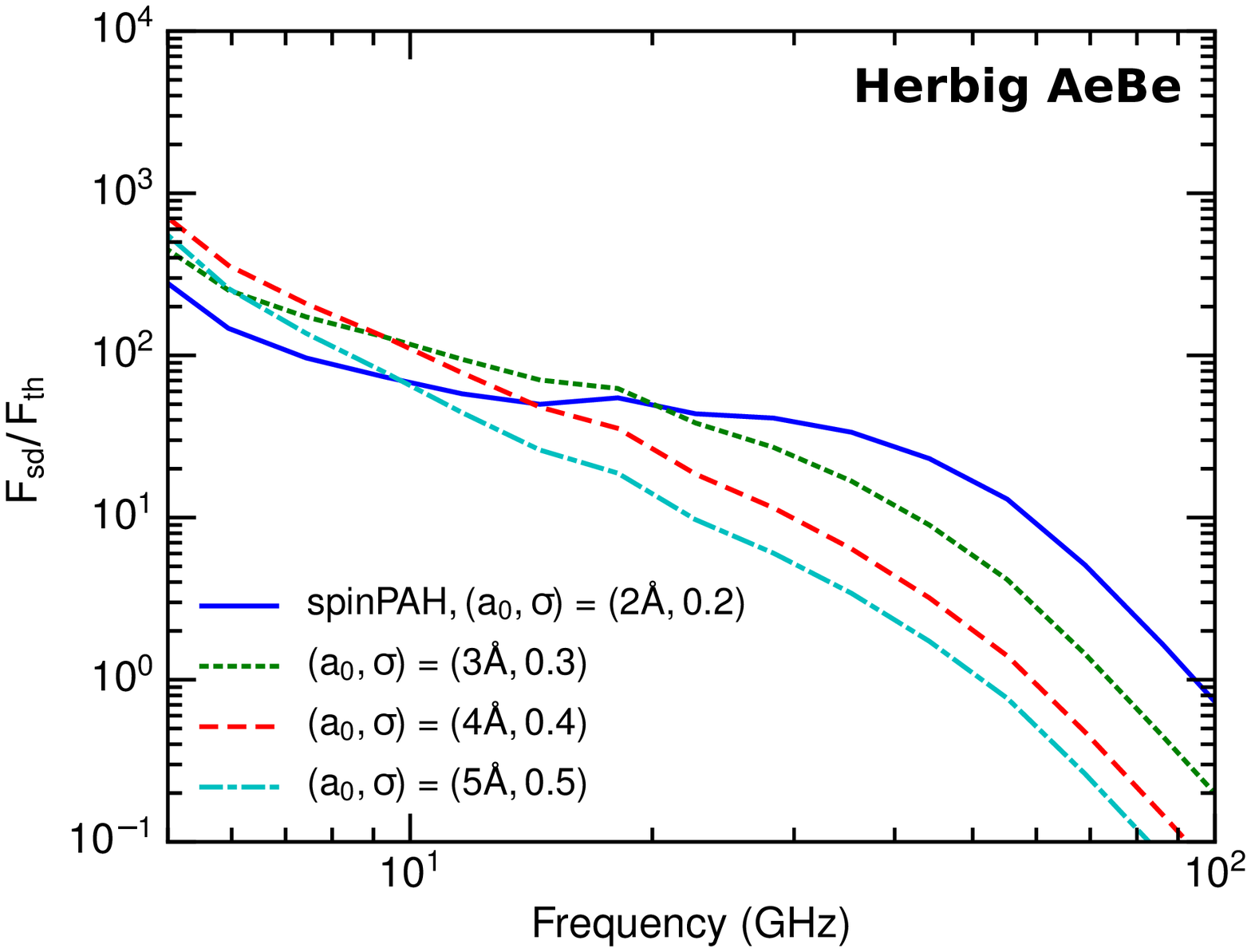}
\includegraphics[width=0.5\textwidth]{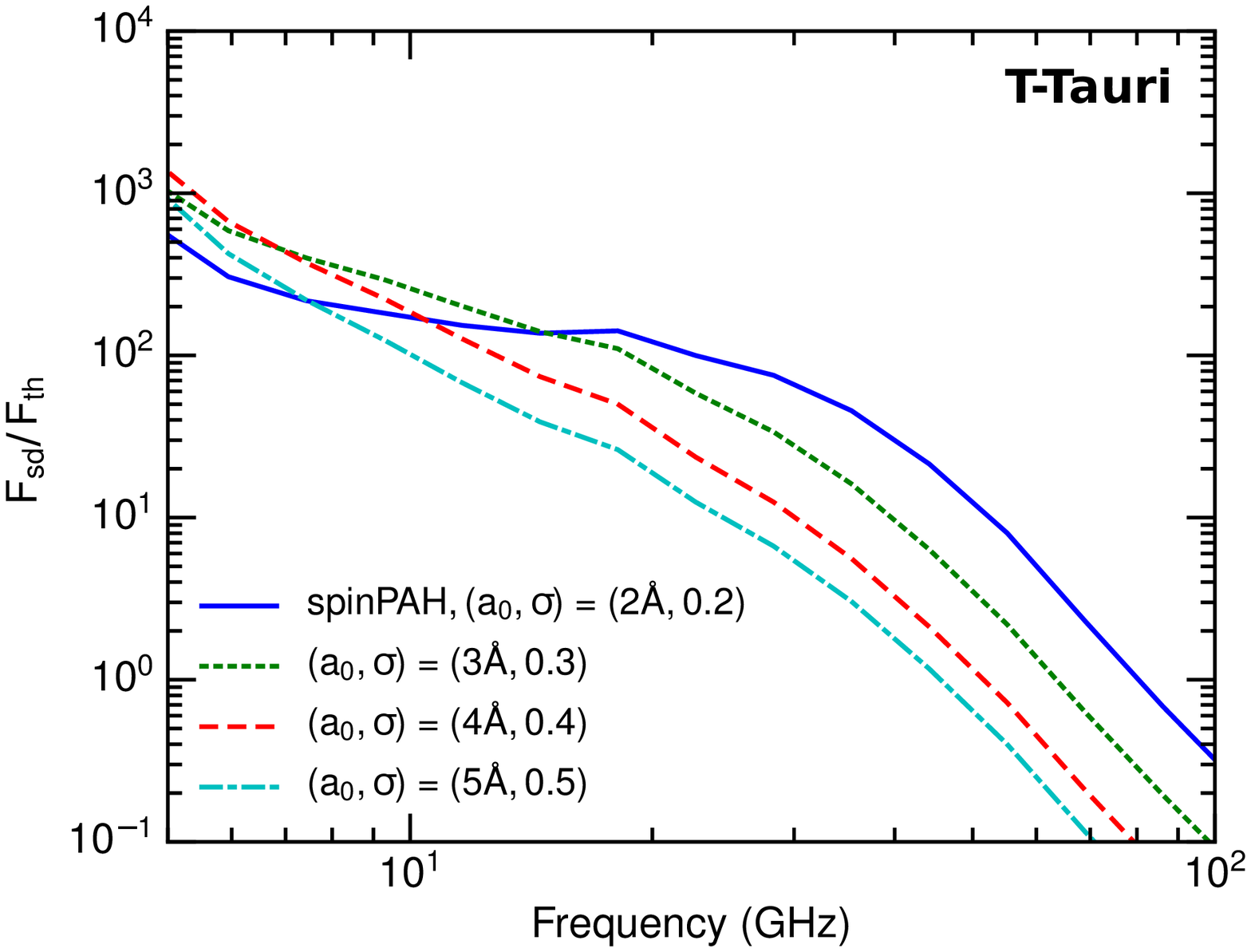}
\caption{Same as Figure \ref{fig:Fspd_disk_PAH}, but for the ratio $F_{\rm sd}/F_{\rm th}$ for the different parameters $a_{0},\sigma$ and fixed $a_{\max}=100$ mm.}
\label{fig:Fspd_Fth_PAH}
\end{figure*}

Figure \ref{fig:Fspd_Fth_PAH} shows the ratio of spectral flux density from spinning PAHs to thermal dust. Emission from spinning PAHs dominates the thermal dust for frequency $\nu<100$ GHz for Herbig Ae/Be disk, and $\nu<40$ GHz for T-Tauri disks.

\subsection{Microwave emission from spinning nanosilicates}
Nanosilicates are expected to have a larger lower cutoff due to more efficient sublimation, as \cite{2017ApJ...836..179H}, thus, we adopt $a_{\rm min}=4.5$\AA. We now vary the Si abundance contained in nanoparticles from $Y_{\rm Si}=0.01$ to $0.2$, while the size distribution parameters are fixed with $a_{0}=3$ \AA~ and $\sigma=0.3$. Indeed, the variation of ($a_{0} ,\sigma$) should produce the similar behavior as in spinning PAH emission because the physics is the same.

\begin{figure*}
\includegraphics[width=0.5\textwidth]{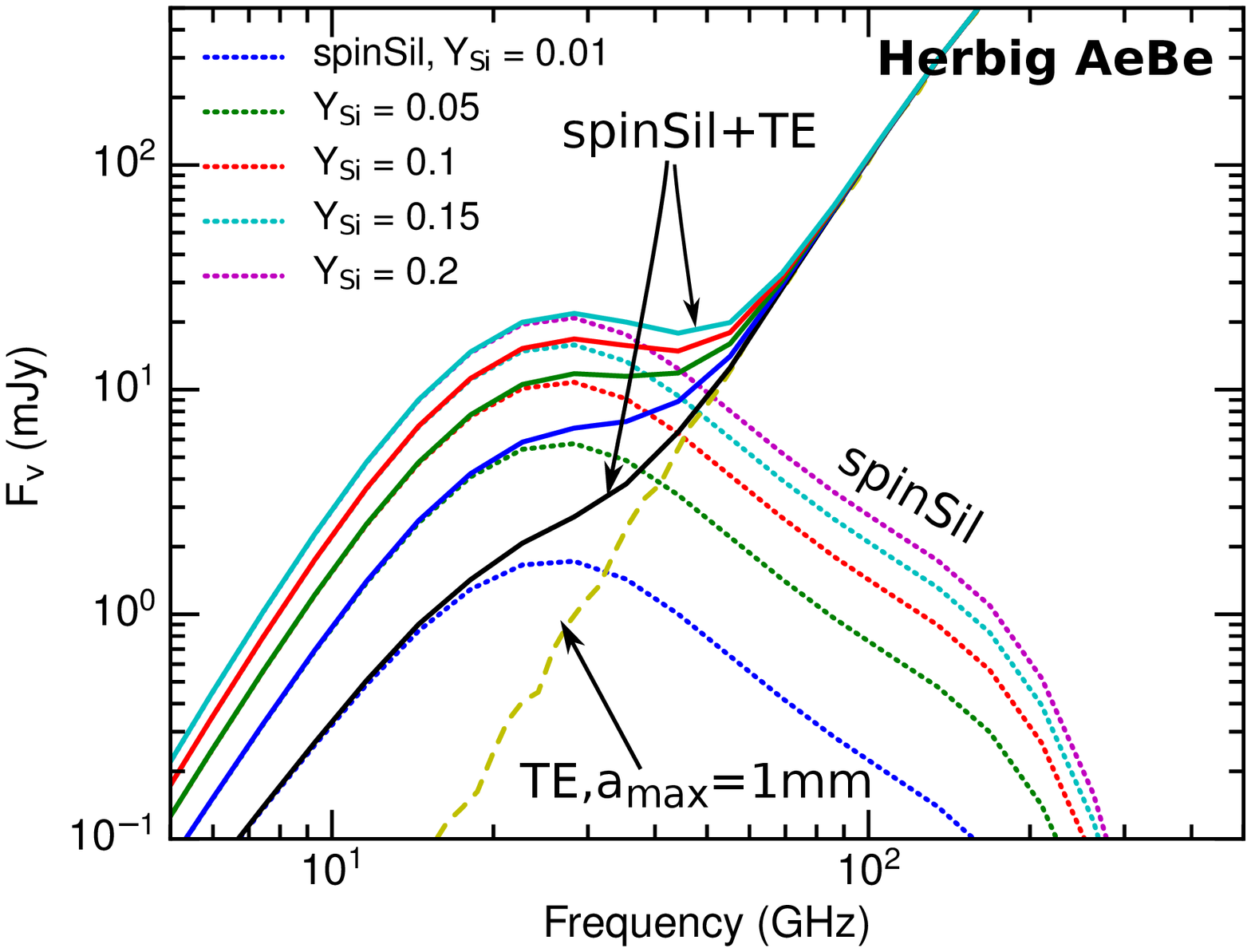}
\includegraphics[width=0.5\textwidth]{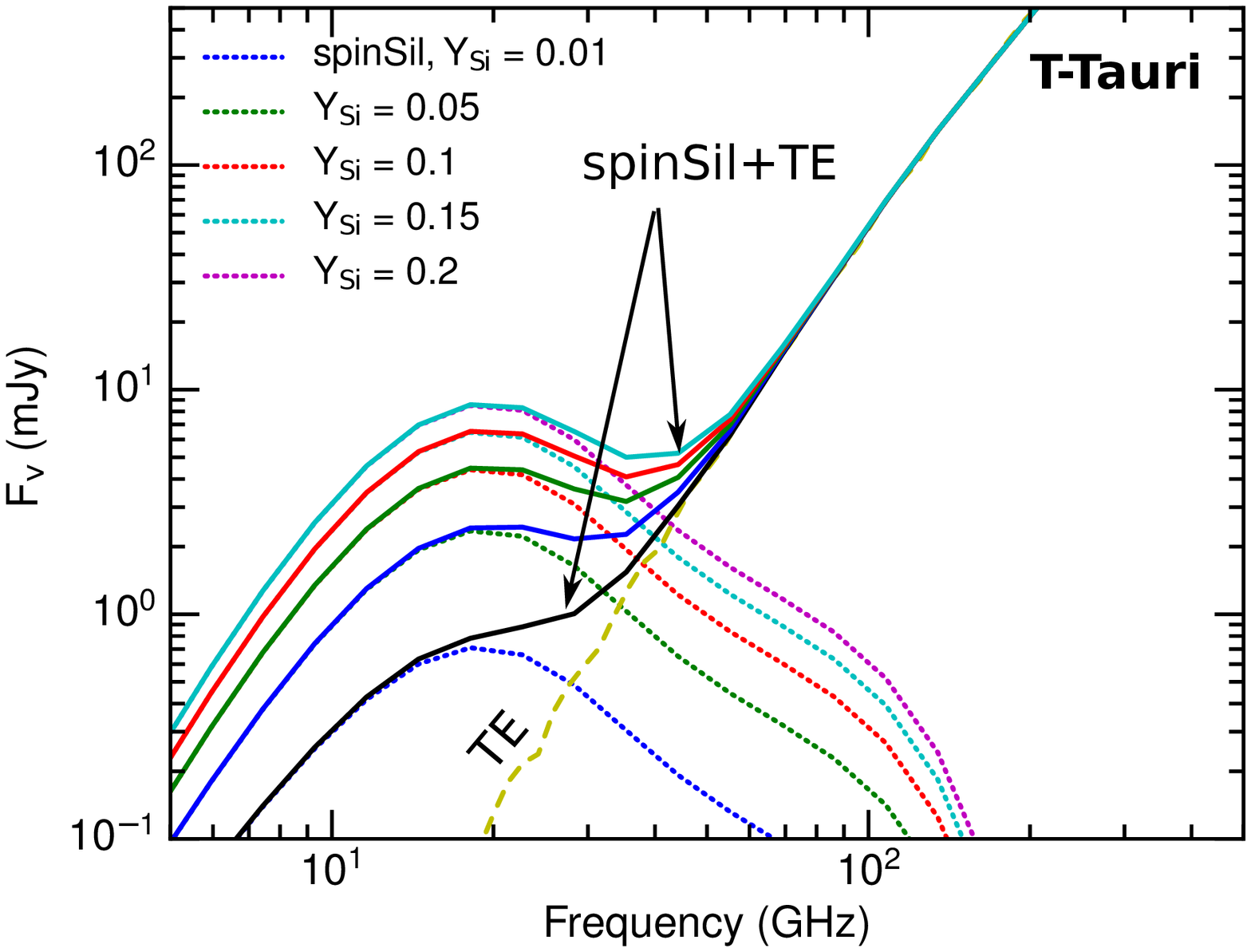}
\caption{Spectral flux density from spinning nanosilicates (dotted lines), thermal dust (dashed line), and the total emission (solid lines). The Si abundance in nanoparticle particles is varied between $Y_{\rm Si}=0.01-0.2$. Left panel and right panel show the results for Herbig AeBe disk and T-Tauri disk, respectively.}
\label{fig:Fspd_disk_sil}
\end{figure*}

In Figure \ref{fig:Fspd_disk_sil}, we plot five spectra of the spinning dust from the disk for the different values of $Y_{\rm Si}$. Emission from spinning nanosilicates is as strong as spinning PAHs for $Y_{\rm Si}=0.05$, as expected, although the maximum emissivity occurs at a lower frequency because of smaller $a_{\rm min}$. Spinning dust flux is increased with increasing $Y_{\rm Si}$. The total emission from spinning PAHs and nanosilicates is much greater than the thermal dust emission at $\nu<100$ GHz. 

\begin{figure*}
\includegraphics[width=0.5\textwidth]{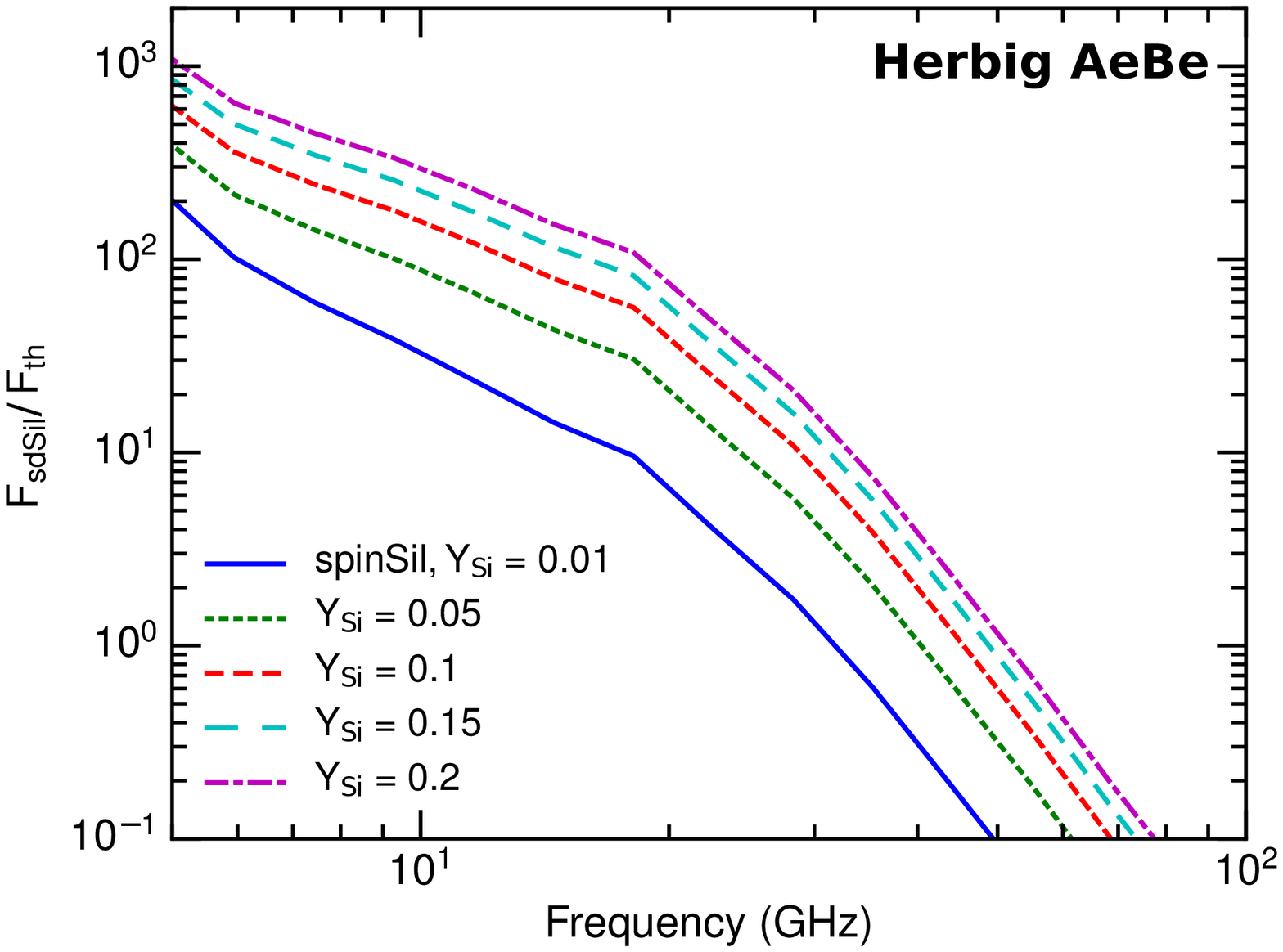}
\includegraphics[width=0.5\textwidth]{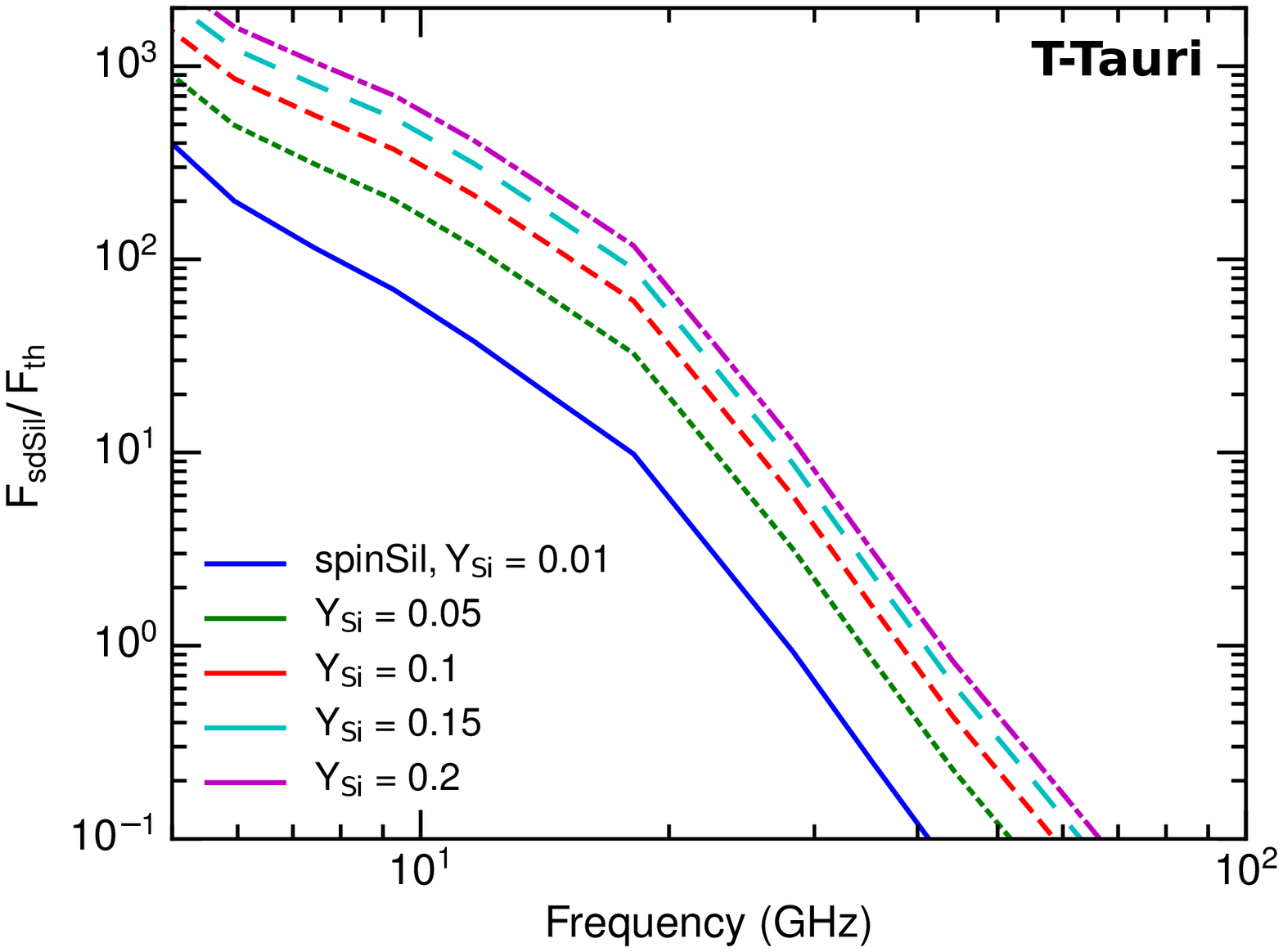}
\caption{Same as Figure \ref{fig:Fspd_disk_sil}, but for the ratio of spectral flux densities from spinning nanosilicates to thermal dust emission for the different $Y_{\rm Si}$.}
\label{fig:Fspd_to_Fth_sil}
\end{figure*}

Figure \ref{fig:Fspd_to_Fth_sil} shows the ratio of spinning dust to thermal dust flux densities. The spinning dust dominates over the thermal dust for $\nu<60$ GHz, even with only 1 percent of Si abundance contained in nanoparticles.

\subsection{Effect of grain growth on mm-cm thermal emission}
\begin{figure*}
\includegraphics[width=0.5\textwidth]{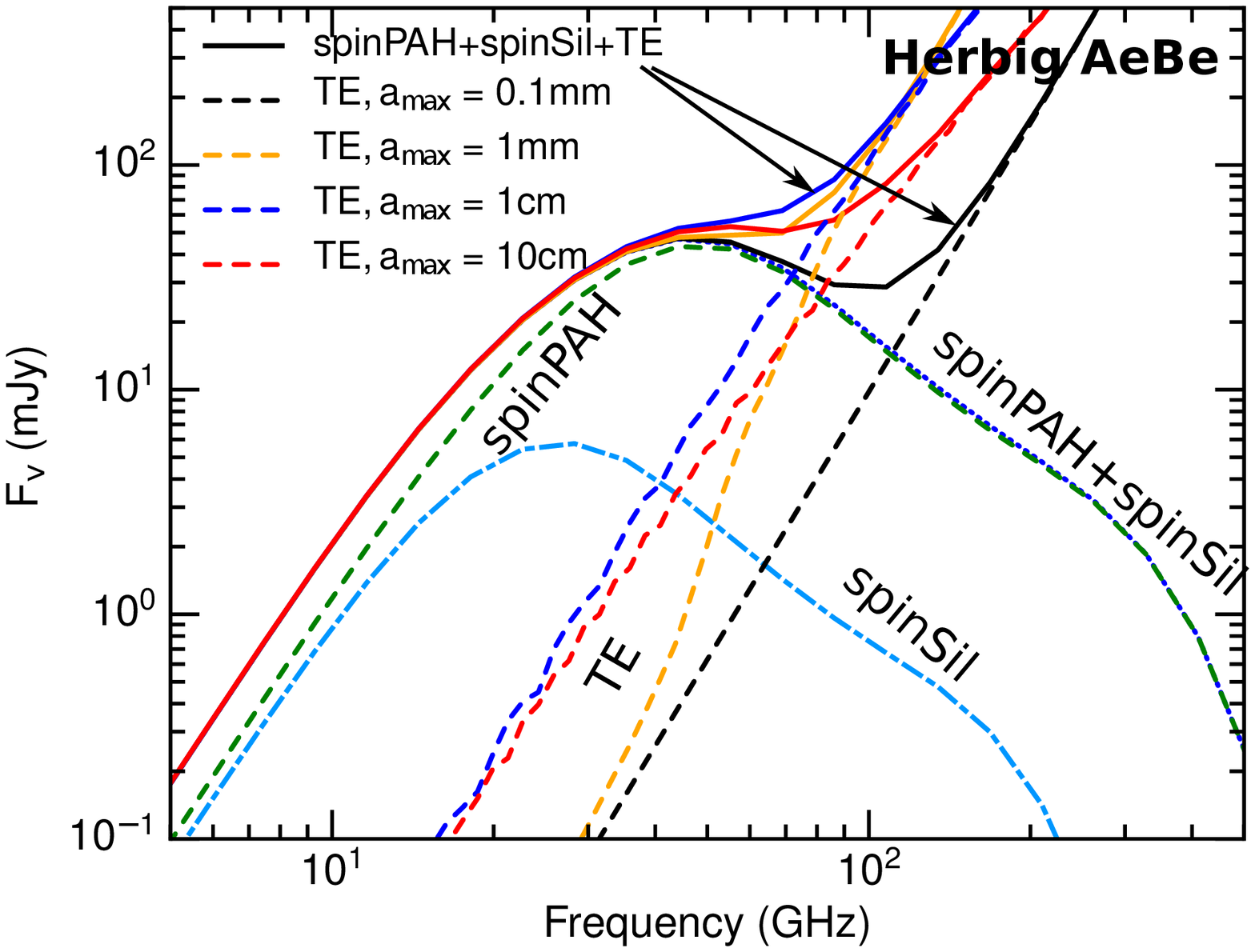}
\includegraphics[width=0.5\textwidth]{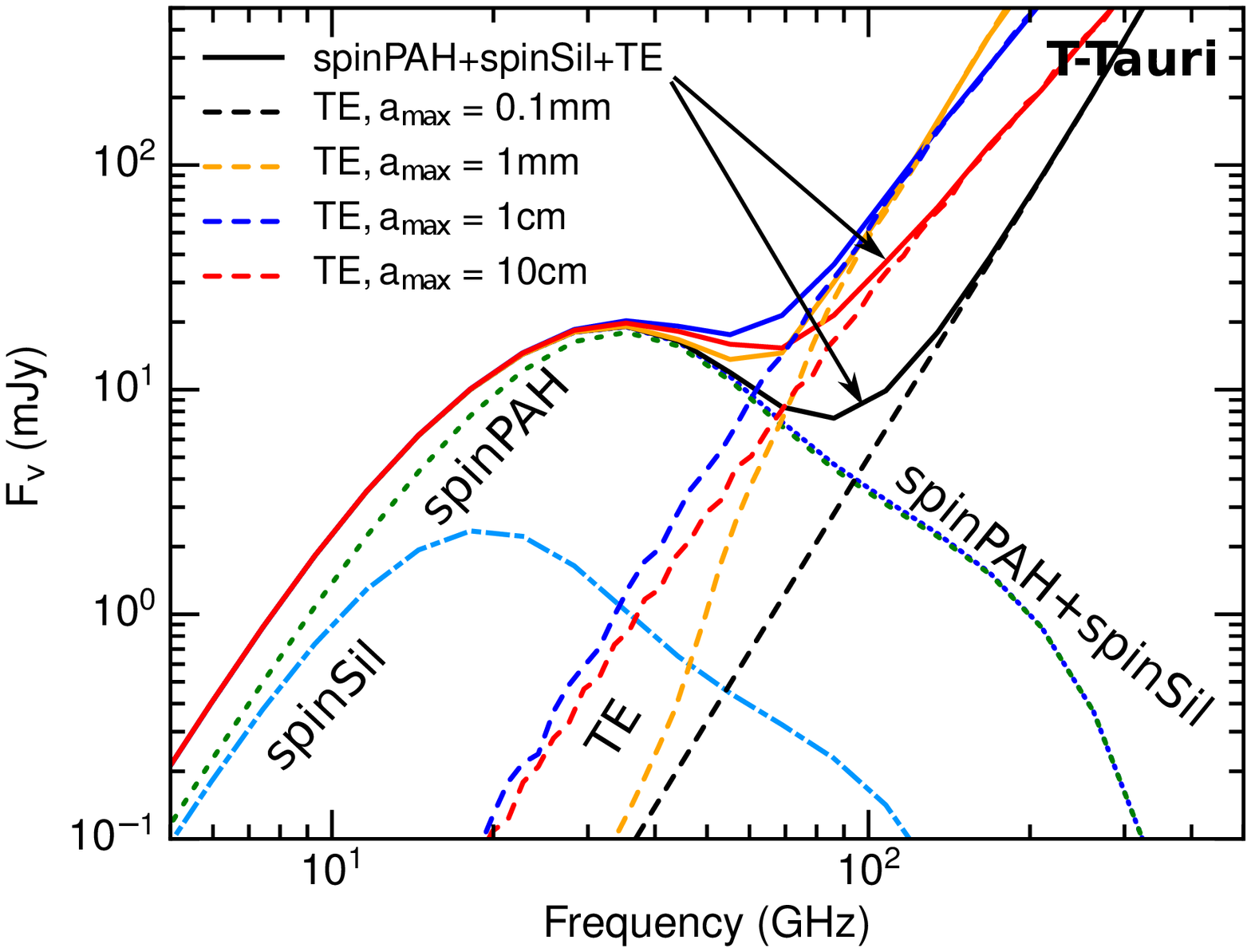}
\caption{Spectral flux density from spinning PAHs (dashed line), spinning nanosilicates (dotted line), and total spinning dust and thermal dust emission (solid lines). Thermal dust emission computed for the different values of $a_{\max}$ are considered, assuming a typical model of spinning PAHs and spinning nanosilicates. Left panel and right panel show the results for Herbig Ae/Be disk and T-Tauri disk, respectively.}
\label{fig:Fspd_disk_PAHsil}
\end{figure*}

To quantify the effect of grain growth on mm-cm thermal emission, in Figure \ref{fig:Fspd_disk_PAHsil}, we show the thermal emission for the different $a_{\max}$ spanning 0.1 mm to 10 cm. The variation of thermal emission is noticeable for $\nu\sim 30-100$ GHz, but the increase in thermal emission from $a_{\max}=5$ mm to $a_{\max}=5$ cm is negligible at $\nu \sim 30-100$ GHz.

\begin{figure*}
\includegraphics[width=0.5\textwidth]{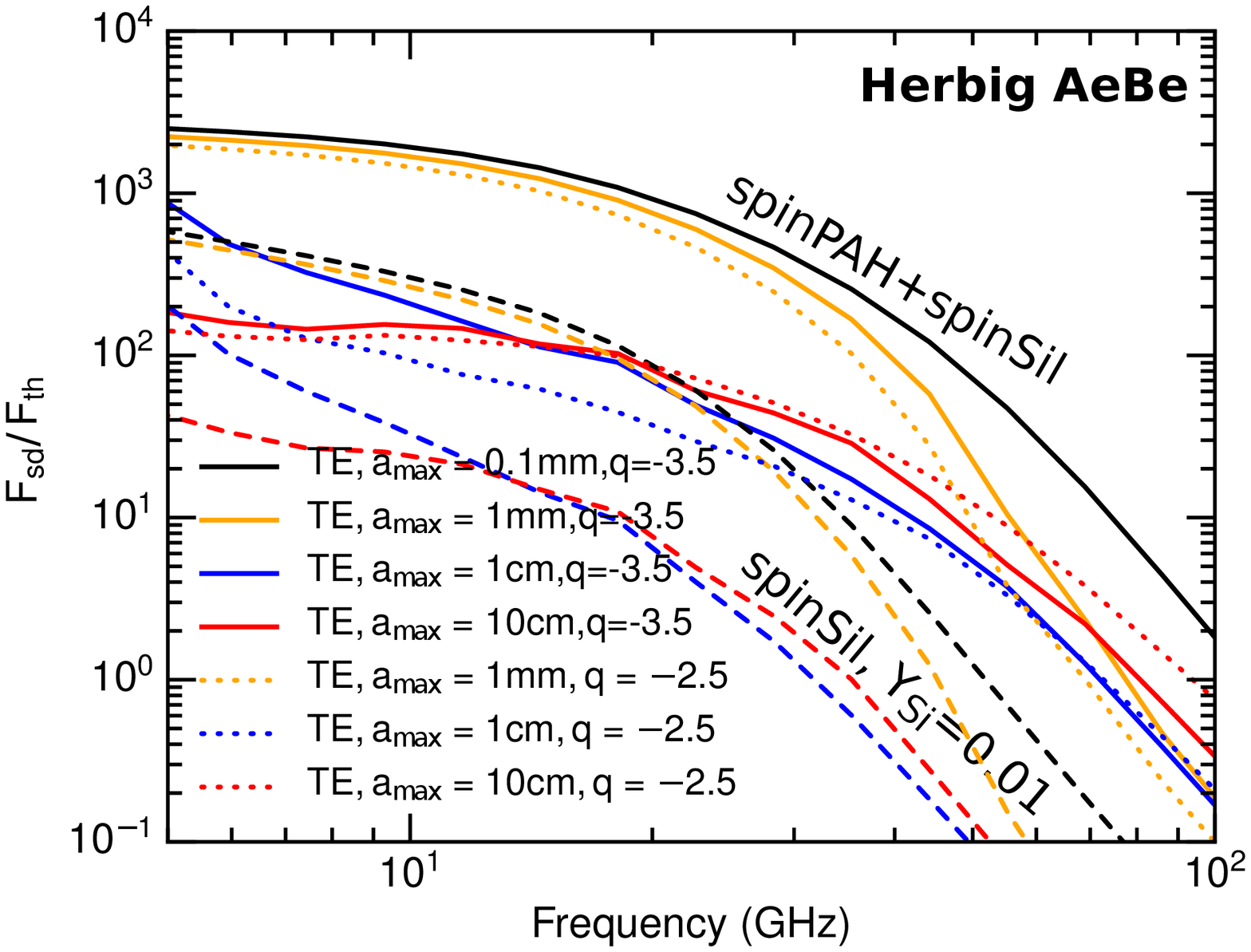}
\includegraphics[width=0.5\textwidth]{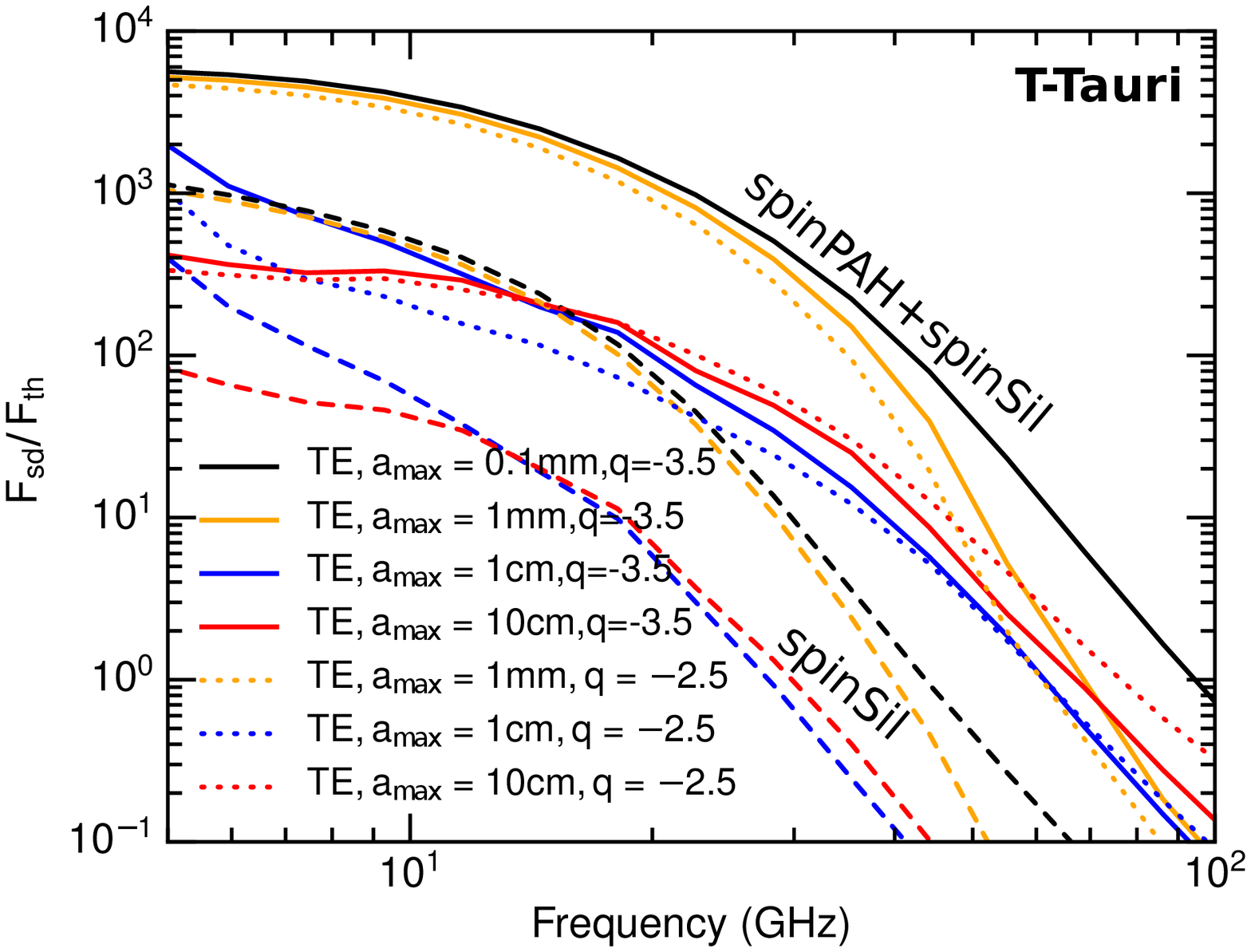}
\caption{Ratio $F_{\rm sd}/F_{\rm th}$ for the different $a_{\max}$. Solid and dotted lines show the results for $q=-3.5$ and $-2.5$, respectively. Dashed lines show the results when only emission from spinning nanosilicates with $Y_{\rm Si}=0.01$ is considered. Left panel and right panel show the results for Herbig Ae/Be disk and T-Tauri disk, respectively.}
\label{fig:Fspd_to_Fth_PAHsil}
\end{figure*}

Figure \ref{fig:Fspd_to_Fth_PAHsil} shows the ratio $F_{\rm sd}/F_{\rm th}$ for the different values of $a_{\max}$. The increase from $a_{\max}=1$ mm to $1$ cm increases the thermal dust emission significantly, resulting in the reduction of $F_{\rm sd}/F_{\rm th}$ by an order of magnitude. The more shallow size distribution helps to enhance thermal dust emission. However, spinning dust is still dominant at frequencies below 60 GHz. The dashed lines show that even only $1\%$ of Si contained in nanosilicates can still produce substantial microwave emission compared to thermal dust with grain growth at $\nu<30$ GHz.

\section{Discussion}\label{sec:discuss}
\subsection{PAHs/Nanoparticles traced by mid-IR emission and implication for spinning dust}
PAH molecules are widely detected in circumstellar disks around Herbig Ae/Be stars and some T-Tauri stars (\citealt{2004A&A...427..179H}; \citealt{2017ApJ...835..291S}). The presence of nanosilicates is also demonstrated by $9.7\mu$m emission features present in many PPDs (\citealt{2017ApJ...835..291S}). 

Recently, modeling works have been done to constrain the physical properties of PAHs. \cite{2003ApJ...594..987L} inferred the PAH size distribution ($a_{0}, \sigma$) by fitting the mid-IR spectrum for the disk around HD 141569A. \cite{2017ApJ...835..291S} derived the PAH size distribution and the total mass of PAHs in about 60 disks around Herbig Ae/Be and T-Tauri stars. The authors found that small PAHs, characterized by $(a_{0},\sigma) \sim (2\AA, 0.2)$, are ubiquitous in PPDs.

If the size distribution of nanoparticles from the shielded region is not different from the surface layer,\footnote{Apparently, the PAH parameters describe PAH molecules from the surface layer directly illuminated by UV radiation. Nevertheless, the vertical mixing is efficient due to turbulence (\citealt{2010A&A...511A...6S}; \citealt{2012A&A...543A..25S}), leading to the frequent circulation of PAHs and nanoparticles between the surface layer and disk interior.} as constrained by mid-IR features, then many disks that have small PAHs inferred in \cite{2017ApJ...835..291S} would provide strong spinning dust emission, provided that C abundance in PAHs $b_{C}>0.01$ (see Figure \ref{fig:Fspd_Fth_PAH}). These disks appear to be the most favorable targets for future observations of spinning dust. 

\subsection{Comparison to previous works}
\cite{2006ApJ...646..288R} carried out a one-dimensional (1D) modeling of microwave emission from spinning PAHs for the disk interior for fiducial disks around Herbig Ae/Be, T-Tauri, and brown dwarf stars. \cite{2006ApJ...646..288R} assumed the thermal rotation (i.e., $T_{\rm rot}=T_{\gas}$) and adopted the standard size distribution of PAHs from the diffuse ISM, with $b_{\rm C}=0.05$. The gas and dust temperature is assumed to follow an analytical formula as a function of the radial distance. Thermal dust is modeled by a power law with a constant spectral slope $\beta$, although the slope $\beta$ at $\nu< 100$ GHz is not a simple function of the maximum grain size $a_{\max}$ (see Figure \ref{fig:kappa_abs}).

In this paper, we have performed self-consistent, two-dimensional (2D) modeling of spinning dust emission (including radial and vertical structures), which is combined with Monte Carlo radiative transfer modeling of thermal dust emission using \textsc{radmc-3d}. In this way, we naturally account for spinning dust emission from both the surface layer and disk interior. We considered a variety of PAH size distributions ($a_{0},\sigma$) that captures the inferred distribution from mid-IR emission (see the preceding section). Moreover, we took into account the emission from rapidly spinning silicate nanoparticles (\citealt{2016ApJ...824...18H}; \citealt{2017ApJ...836..179H}). We found that microwave emission from nanosilicates could significantly increase the SED at $\nu<60$ GHz, making the detection more easy than spinning PAHs alone. Previous studies by \cite{2016ApJ...824...18H} show that Si abundance $Y_{\rm Si}$ can reach $10\%$ without violating observational constraints in UV extinction, AME polarization, and IR emission. We find that even $Y_{\rm Si}=0.01$ can produce the AME by a factor of 10 larger than the thermal dust for $\nu\sim 30$ GHz.

In particular, the flux density of thermal dust emission from a circumstellar disk is calculated by \textsc{radmc-3d} for the different dust size distributions in the presence of grain growth with $a_{\max}$ spanning from 0.1 mm to 10 cm. The simultaneous modeling of spinning dust and thermal dust with grain growth allows us to quantify the respective contribution to microwave emission by these two mechanisms as a function of the nanoparticle size distribution and maximum values $a_{\max}$. 

\subsection{Can spinning dust explain excess microwave emission from circumstellar disks?}
\subsubsection{Excess microwave emission (EME) from disks}
EME is often found in radio observations from circumstellar disks around Herbig Ae/Be stars (\citealt{1993ApJS...87..217S}; \citealt{2001A&A...365..476M}; \citealt{2006MNRAS.365.1283D}; \citealt{2011ApJ...727...26S}; \citealt{vanderPlas:2016gx}), as well as T-Tauri stars (\citealt{2002ApJ...568.1008C}; \citealt{2004A&A...416..179N}; \citealt{2005ApJ...626L.109W}; \citealt{2012MNRAS.425.3137U}). The popular explanations for such EME include thermal dust emission from cm-sized grains and free-free emission from winds/ jets (see e.g., \citealt{2012MNRAS.425.3137U}). Recently, \cite{2017MNRAS.466.4083U} observed the emission excess from 11 disks around T-Tauri stars and suggested that multiple mechanisms should be responsible for EME.

\subsubsection{Spinning dust as an origin of EME}
Investigating observational data collected from the literature presented in \citealt{2011ApJ...727...26S}, one can see that the Herbig Ae/Be disks with prominent EME include R Mon, HD 35187, HD 163296, HD 169142. Interestingly, the three latter disks also exhibit prominent PAH emission (see \citealt{2017ApJ...835..291S}), while weak PAH emission is observed in the R Mon disk \citep{2012A&A...538A.101V}. The HD 35187 and HD 163296 disks also exhibit strong 9.7 $\mu$m silicate emission. Thus, we expect some contribution of spinning dust to the observed EME.

To explore whether spinning dust can explain EME from circumstellar disks, we first fit the observational data with a two-component model, including thermal dust and spinning dust. The total flux density is described by:
\bea
F_{\rm mod}(\nu)& = &~F_{\rm td, 100}\left(\frac{\nu}{100\GHz}\right)^{\alpha_{td}}\nonumber\\
&& + F_{\rm sd,0}\left(\frac{\nu}{\nu_{\rm pk}}\right)^{2}\exp\left[1-\left(\frac{\nu}{\nu_{pk}}\right)^{2}\right],\label{eq:Imod}
\ena
where $\nu_{\rm pk}$ and $F_{\rm sd,0}$ are the frequency and the flux density at the peak of the spinning dust spectrum (see \citealt{2012ApJ...757..103D}), $F_{\rm td,100}$ is the thermal emission flux density measured at 100 GHz, and $\alpha_{td}=\beta+2$ with $\beta$ the spectral slope of the dust opacity. The model parameters include $F_{\rm td, 100}, \beta, F_{\rm sd,0}$, and $\nu_{\rm pk}$. \footnote{Due to the limited observational data at $\nu<100$ GHz, we fit with a parametric model instead of performing physical modeling of spinning dust because the physical model depends on $\sim$ 10 parameters (e.g., dipole moment, size distribution, and gas density and temperature). A three-component fitting, including thermal dust, spinning dust, and free-free emission is also not feasible due to the same reason.}

\begin{figure*}
\includegraphics[width=0.5\textwidth]{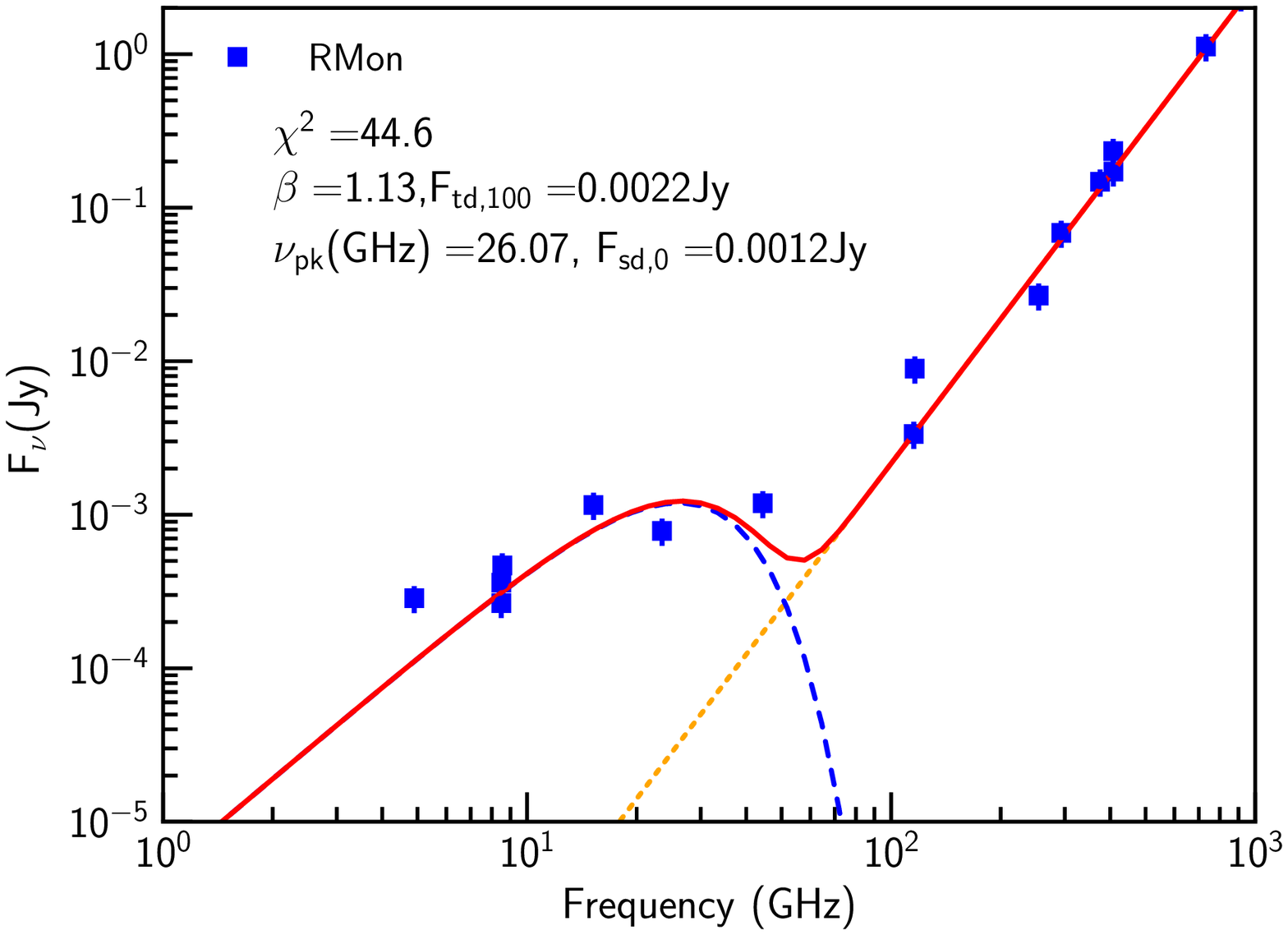}
\includegraphics[width=0.5\textwidth]{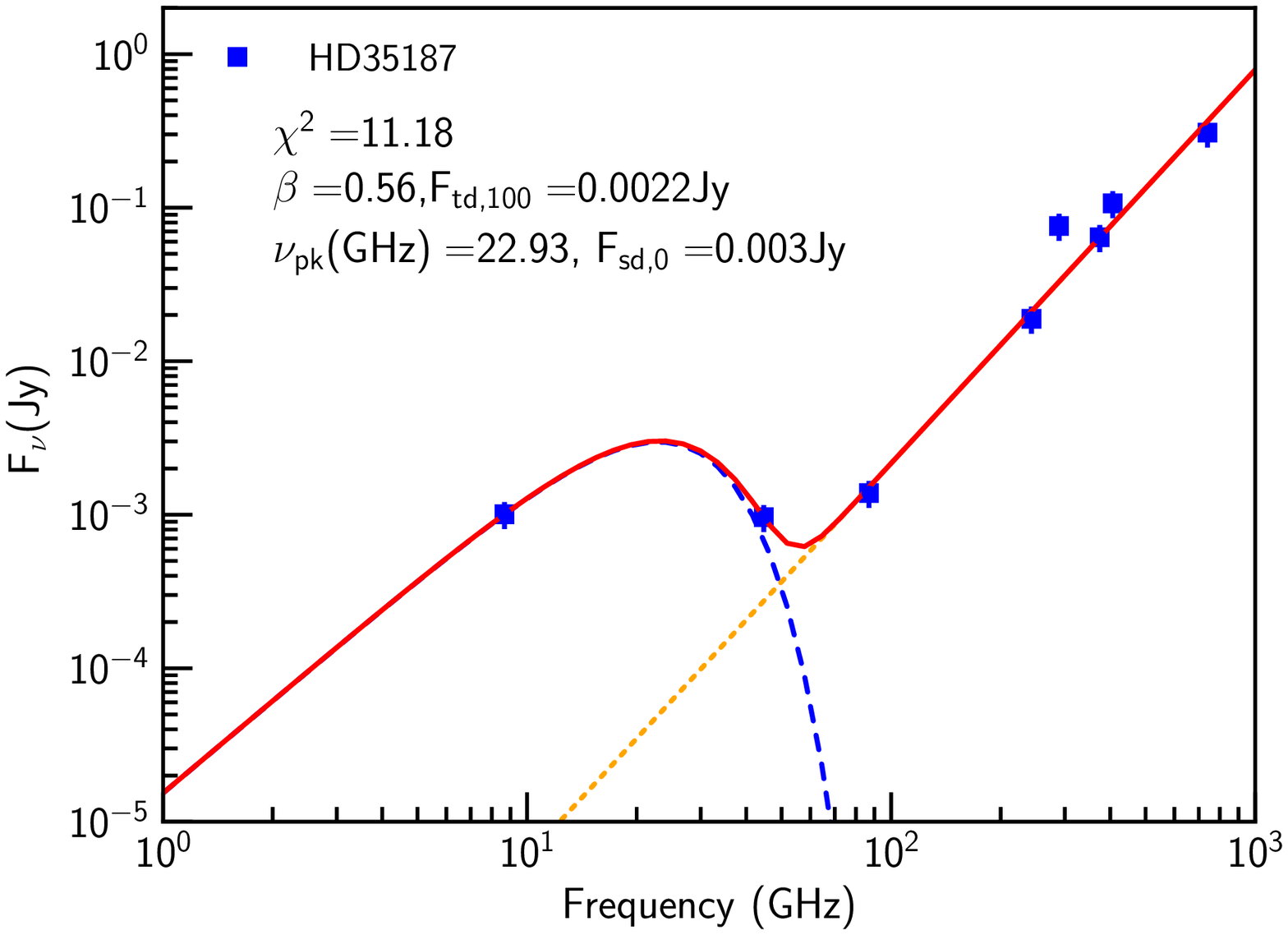}
\includegraphics[width=0.5\textwidth]{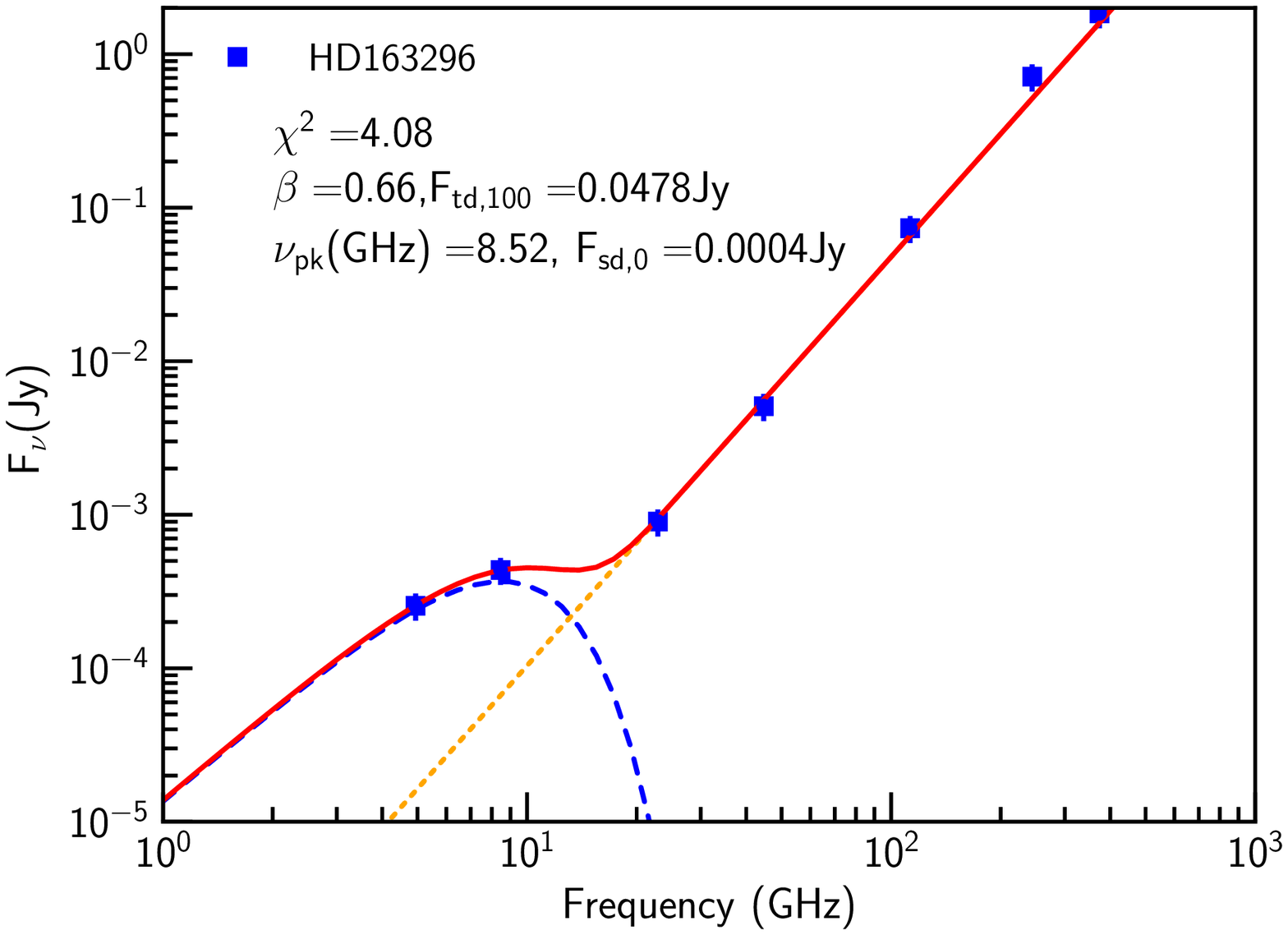}
\includegraphics[width=0.5\textwidth]{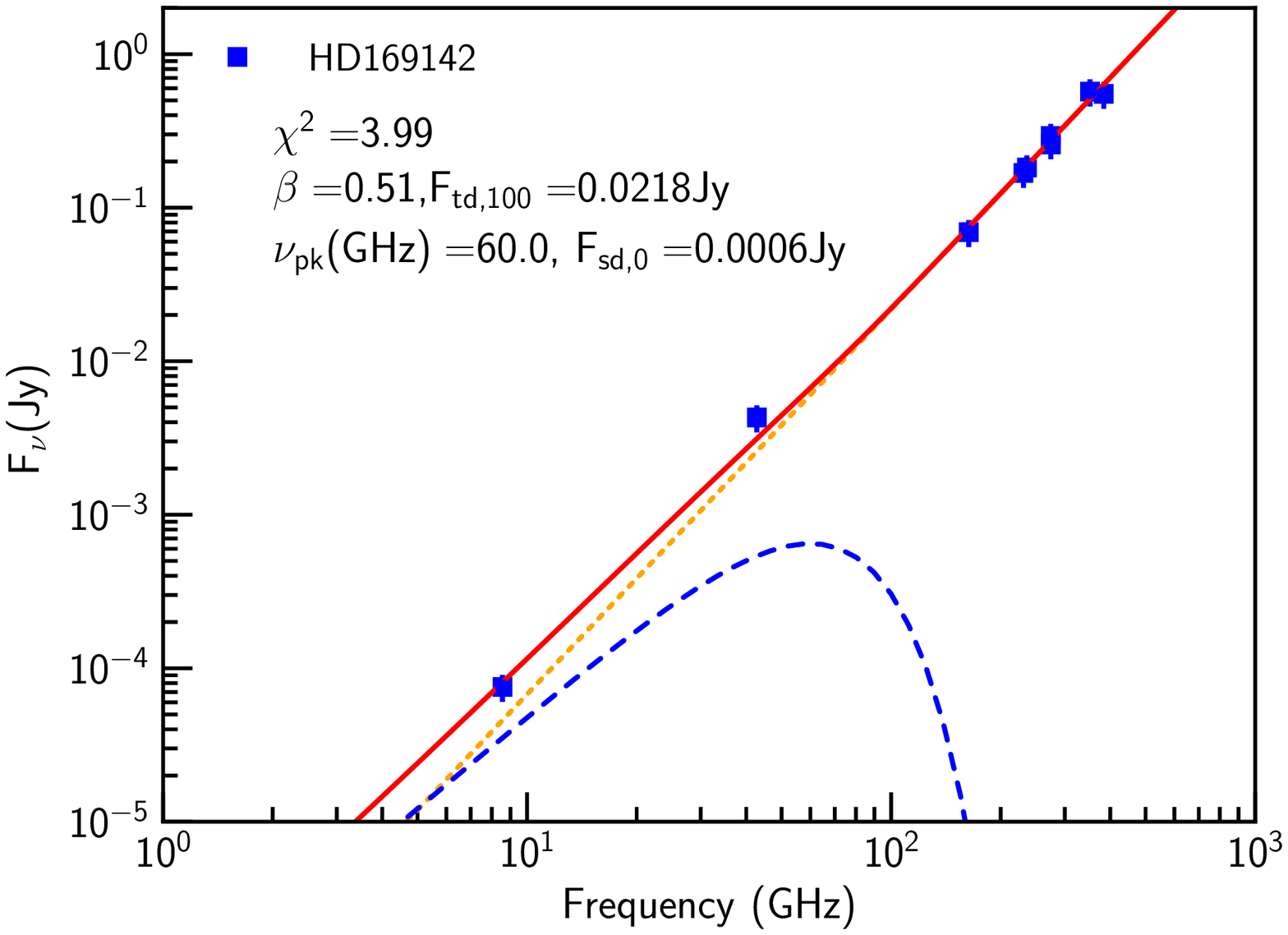}
\caption{Two-component fitting, spinning dust (blue line) and thermal dust (orange line) to the mm-cm data (squared symbols) for the disks around Herbig Ae/Be stars. Best-fit model parameters and the corresponding $\chi^{2}$ are indicated. The total best-fit model is shown in red line.}
\label{fig:SED_fit}
\end{figure*}

The goodness of fit to the observed flux, $F_{\rm obs}(\nu)$, is measured by $\chi^{2}$, as defined by
\bea
\chi^{2}=\sum_{j}\left(\frac{F_{\rm mod}(\nu_{j}) - F_{\rm obs}(\nu_{j})}{\sigma_{j}}\right)^{2},
\ena
where $\sigma_{j}$ is the data uncertainty at the data point $\nu_{j}$, which is fixed to be $20\%$ of $F_{\rm obs}(\nu_{j})$. We infer the best-fit model parameters by minimizing $\chi^{2}$ using the Levenberg-Marquart method from a publicly available package \textsc{lmfit} \citep{newville_2014_11813}. We fit to the mm-cm (i.e., $\nu < 1000$ GHz) data obtained from \cite{2011ApJ...727...26S} and \cite{2001A&A...365..476M}.

In Figure \ref{fig:SED_fit}, we show our two-component fits to the observational data for four disks around Herbig Ae/Be stars. For R Mon and HD 163296 disks, the required spinning dust flux is $F_{\rm sd,0}\sim 0.5-1$ mJy, which can easily be reproduced with the low PAH/Si abundance of $Y=0.01$ (see Figure \ref{fig:Fspd_disk_sil}). To reproduce the data for HD 163296, it requires spinning dust to peak at $\nu_{\rm pk}\sim 8.5$ GHz, while HD 169142 requires $\nu_{\rm pk}\sim 60$ GHz. Incidentally, mid-IR modeling by \cite{2017ApJ...835..291S} find that the HD 169142 disk contains small PAHs (i.e., $a_{0},\sigma=(2.5\AA, 0.2)$), while the HD 163296 disk contains larger PAHs (i.e., $a_{0},\sigma=(4.0\AA, 0.2)$). Such small/large PAHs are predicted to strongly emit microwave emission with a high/low peak frequency (see Figure \ref{fig:Fspd_disk_PAH}), consistent with the peak frequencies inferred from the model fitting.

Our best-fit thermal dust yields $\beta\sim 0.5-1.2$, implying the presence of cm-sized grains in these disks (see \citealt{2006ApJ...636.1114D}). The study of pebbles and planetesimals in PPDs using ALMA Band 1-3 and SKA (see e.g., \citealt{2015aska.confE.117T}) would suffer contamination from spinning dust at $\nu < 100$ GHz. Therefore, to achieve a realistic measurement of the thermal dust spectral slope $\beta$ and realistic understanding of planet formation, spinning dust needs to be carefully modeled and separated from the observational data. 

\subsubsection{On the importance of free-free emission}
{At microwave frequencies, free-free emission from stellar winds or ionized jets is expected to be important in circumstellar disks. Its emission flux can be described by a power law, $F_{\rm ff}\propto \nu^{\alpha_{\rm ff}}$, where $\alpha_{\rm ff}$ is the spectral slope. For optically thin region, $\alpha_{\rm ff}=-0.1$, but $\alpha_{\rm ff}$ becomes positive and can reach $\alpha_{\rm ff}\sim 2$ for optically thick regions (\citealt{1986ApJ...304..713R}). With this wide range of values, free-free emission is a leading mechanism to explain the EME (cf., see \citealt{2017MNRAS.466.4083U}). Here, we have also attempted to fit the EME with a model consisting of free-free emission and thermal dust emission. As expected, free-free emission can provide an equally good fit to the observational data as spinning dust. Specifically, the best-fit spectral index is $\alpha_{\rm ff}\sim 0.7$ for R Mon, $\sim 1$ for HD 164192, -0.02 for HD 163296, and -0.1  for HD 35187.

Finally, we should stress that, except R Mon and HD 163296, two other disks (HD 35187 and HD 169142) have insufficient data points below 100 GHz to allow a robust constraint on the actual role of spinning dust for EME. Future multi-frequency observations between 1-60 GHz by SKA, ngVLA, and ALMA Band 1 and 2 \citep{Fuller:2016um} would be valuable to differentiate spinning dust and free-free emission as an origin of EME in circumstellar disks. Moreover, polarization observations would be particularly useful because free-free emission is unpolarized. It also can constrain the carriers of AME because the polarization of spinning nanosilicate emission is expected to be higher than spinning PAHs \citep{2016ApJ...821...91H}.
}

\subsection{Towards probing nanoparticles in circumstellar disks via spinning dust}
PAHs and nanoparticles are expected to play an important role in gas heating, chemistry and dynamics of disks because they contribute the largest surface area for charge carrier and astrochemical activities (see \citealt{2013ApJ...766....8A}). Indeed, nanoparticles characterize the ionization level of the disk interior, which affects the magnetohydrodynamic instability activity and the dead zones (\citealt{2003ApJ...585..908F}). The probe of PAHs through mid-IR emission is limited mostly to the surface region where PAHs are directly exposed to the stellar UV radiation. Therefore, the detection of spinning dust emission in PPDs is not only a smoking-gun for the PAHs as a carrier of AME, but it also provides a new diagnostic for nanoparticles in the entire volume of PPDs. 

The non-detection of AME from the disk with prominent PAH features but no silicate emission features would provide a convincing test for spinning PAHs as a carrier of the AME. Similarly, the detection/non-detection of AME from the disks with silicate features would provide a valuable test for the spinning nanosilicates as a carrier of AME. 

It is worth to mention that (sub)mm-wavelength observations usually reveal central cavities and gaps in transitional disks (e.g., HD 169142 \citealt{2017A&A...600A..72F}). This indicates that significant grain growth has occurred so that its thermal emission is substantially reduced in (sub)mm wavelengths. If the assumption of PAHs/VSGs mixed to the gas is valid, then, we expect to detect spinning dust emission by these nanoparticles from cavities and gaps. Therefore, transitional disks appear to be excellent target to study spinning dust with future high-resolution experiments like ALMA Band 1, ngVLA, SKA. Interestingly, a marginal detection of 33 GHz signal from the intracavity in MWC 758 is recently reported by \cite{Casassus:2018td}, which is suggested to be spinning dust emission. 

\section{Summary}\label{sec:sum}
We studied microwave emission from rapidly spinning nanoparticles from circumstellar disks around Herbig Ae/Be stars and applied to explain the observed excess microwave emission. The principal results are summarized as follows:

\begin{itemize}

\item[1]
We performed a physical, two-dimensional modeling of microwave emission from both rapidly spinning PAHs and spinning nanosilicates in circumstellar disks that include both for the disk interior and surface layers. The dust temperature is numerically computed using the Monte Carlo radiative transfer code (\textsc{radmc-3d}).

\item[2] 
We found that microwave emission from either spinning PAHs or spinning nanosilicates can dominate over thermal dust at frequencies $\nu<60$ GHz in circumstellar disks. The presence of both spinning nanosilicates and PAHs can significantly increase the spectral flux density at $\nu <100$ GHz. Our obtained results imply that the possibility to detect spinning dust emission in PPDs is much higher than previously thought.

\item[3] 
By simultaneous modeling of spinning dust and thermal dust emission for a physical disk model, we showed that the thermal dust is still much lower than spinning dust at $\nu<60$ GHz, even the maximum grain size is increased 10 cm. The presence of spinning dust emission would complicate the probe of grain growth and formation of planetesimals using radio observations. 

\item[4] Our two-component (thermal dust and spinning dust) model fitting to the mm-cm observational data of several Herbig Ae/Be disks (R Mon, HD 163296, HD 35187, and HD 169142) reveal that spinning dust can reproduce excess microwave emission from the disks. Future multi-frequency observations by ALMA, ngVLA, and SKA would be valuable for elucidating the origin of EME as well as AME. Polarization observations would help to distinguish the carriers (PAHs or nanosilicates) of AME. Detection of spinning dust emission in circumstellar disks would open a powerful way to probe nanoparticles and understand its role on disk astrochemistry.

\end{itemize}
\acknowledgments
{We thank the anonymous referee for helpful comments that improve the presentation of this paper.} This work was supported by the Basic Science Research Program through the National Research Foundation of Korea (NRF), funded by the Ministry of Education (2017R1D1A1B03035359). We thank Attila Juhasz for help with radmc3dPy and J-Y Seok for discussion. One of the authors (Quynh Lan N.) was supported by the U.S. Department of Energy under Nuclear Theory Grant DE-FG02-95-ER40934 and in part by the National Science Foundation under Grant No. PHY-1430152 (JINA Center for the Evolution of the Elements).

\appendix

\section{Review of Circumstellar Disk Physics}\label{apdx:diskmod}

\subsection{Stellar radiation}
The surface layer has energy density given by
\bea
u_{\rm rad}(r) = \left(\frac{L_{\star}}{4\pi r^{2}c} \right)^{2},~ \chi=\frac{u_{\rm rad}}{u_{\rm MMP}}\sim 5.3\times 10^{3}\left(\frac{L_{\star}}{L_{\odot}}\right)\left(\frac{100 \AU}{r}\right)^{2},\label{eq:U}
\ena
where with $u_{\rm MMP}=8.64\times 10^{-13}\erg \s^{-1}$ is the typical energy density of the diffuse interstellar radiation from \cite{1983A&A...128..212M}.

\subsection{Disk mass and PAH mass}\label{apd:MPAH}

The total gas and dust mass of a disk is estimated as
\bea
M_{\rm disk}= \int_{R_{\rm in}}^{R_{\rm out}} dr\int dz(2\pi r \mu m_{\H}n_{\H}(r,z)= \int_{R_{\rm in}}^{R_{\rm out}} 2\pi r \Sigma(r)dr=\frac{2\pi \Sigma_{1}AU^{\alpha}}{2-\alpha}\left(R_{\rm out}^{2-\alpha}-R_{\rm in}^{2-\alpha}\right)
\ena
where $\Sigma(r)=\Sigma_{1}(r/\AU)^{-\alpha}$ has been used. For a fiducial disk of $R_{\rm in}=1\AU, R_{\rm out}=300\AU$, and $\Sigma_{1}=1$, we get $M_{\rm disk}\sim 0.2M_{\odot}$, assuming $\alpha=1$. The dust disk mass is $10^{-2}M_{\rm disk}=0.002M_{\odot}$.


The total mass of X nanoparticles (PAHs or nanosil) from both disk and surface layer is evaluated as
\bea
M_{X}&=&\int_{R_{\rm in}}^{R_{\rm out}} dr\int dz(2\pi r n_{\H})\int \frac{dm_{X}}{n_{\H}da} da=\int_{R_{\rm in}}^{R_{\rm out}} 2\pi r \frac{\Sigma(r)}{\mu m_{\H}} dr\int \frac{dm_{X}}{n_{\H}da} da=\int_{R_{\rm in}}^{R_{\rm out}} 2\pi r \frac{\Sigma_{1}}{\mu m_{\H}}r^{-\alpha} dr\int \frac{\rho 4\pi a^{3}dn}{3n_{\H}da} da,\nonumber\\
&=&\frac{M_{\rm disk}}{\mu m_{\H}}m_{X}b_{X}=\frac{M_{\rm disk}}{\mu m_{\H}}\frac{M_{\rm PAH}}{\H}=\frac{M_{\rm disk}}{\mu m_{\H}}m_{X}b_{X},\label{eq:MPAH}
\ena
where $m_{X}$ is the average atomic mass of PAHs, $b_{X}$ is the abundance of C in nanoparticles. For graphene of purely carbon, $m_{X}=m_{\rm C}$.

\subsection{Thermal dust emission}\label{apd:TE}
In addition to spinning emission, the grains thermally heated (by starlight, etc.) in the disk emit thermal emission. The luminosity of emission from the entire disk is equal to
\bea
L_{\nu}=4\pi \nu \int dr (2\pi r)\int dz \alpha_{\nu}B_{\nu}(T_{i})e^{-\tau_{\nu}}=4\pi \nu \int dr (2\pi r)\int d\tau_{\nu}B_{\nu}(T_{i})e^{-\tau_{\nu}},
\ena
where $\alpha_{\nu}$ is the absorption coefficient, $d\tau_{\nu}=\alpha_{\nu} dz=\kappa_{\nu}\rho_{d}(r,z) dz$ is the optical depth along z-direction, and $\tau_{\nu}$ measures the optical depth from $z$ to the infinity (\citealt{2001ApJ...547.1077C}). For an isothermal disk, this integral yields
\bea
F_{\rm th}(\nu)=  \frac{1}{4\pi D^{2}}\int_{R_{\rm in}}^{R_{\rm out}}4\pi B_{\nu}(T_{i})(1-e^{-\tau_{\nu}})2\pi rdr=\frac{1}{4\pi D^{2}}\int_{R_{\rm in}}^{R_{\rm out}}4\pi B_{\nu}(T_{i})(1-e^{-\kappa_{\nu}\Sigma_{d}(r)})2\pi rdr,\label{eq:Fth1}
\ena
where $\tau_{\nu}=0$ at the far-side surface layer.


\bibliography{ms.bbl}

\end{document}